\documentclass[11pt,draftcls]{IEEEtran}
\onecolumn

\usepackage[T1]{fontenc}
\usepackage[mathcal]{euscript}
\usepackage[latin1]{inputenc}
\usepackage{amsmath,amstext,amsfonts,amssymb,amscd}
\usepackage[dvips]{graphicx}
\DeclareGraphicsExtensions{.eps}
\usepackage[tight,footnotesize]{subfigure}
\usepackage{enumerate}

\newcommand{\bs}{\boldsymbol}
\newcommand{\tnorm[1]}{\left|\!\left|\!\left|#1\right|\!\right|\!\right|_\infty}

\makeatletter
\def\overparenthesis#1{\mathop{\vbox{\ialign{##\crcr\noalign{\kern3\p@}
\downparenthfill\crcr\noalign{\kern3\p@\nointerlineskip}
$\hfil\displaystyle{#1}\hfil$\crcr}}}\limits}
\def\downparenthfill{$\m@th\braceld\leaders\vrule\hfill\bracerd$}
\makeatother

\newtheorem{lemma}{Lemma}
\newtheorem{corollary}{Corollary}
\newtheorem{proposition}{Proposition}
\newtheorem{theorem}{Theorem}
\newtheorem{definition}{Definition}
\newtheorem{remark}{Remark}

\def \A {\mathbf{A}}
\def \B {\mathbf{B}}
\def \C {\mathbf{C}}
\def \Csup {C_{\mathrm{sup}}} \def \Cmax {C_{\mathrm{max}}}

\def \H {\mathbf{H}}
\def \Hk {\mathbf{H}^{(k)}} \def \Hkc {\mathbf{H}^{(k)*}} \def \Hl {\mathbf{H}^{(l)}} \def \Hlc {\mathbf{H}^{(l)*}}
\def \I {\mathbf{I}} 
\def \M {\mathbf{M}}
\def \N {\mathbf{N}}
\def \P {\mathbf{P}}
\def \Q {\mathbf{Q}}
\def \R {\mathbf{R}}
\def \Rt {\tilde{\R}}   \def \Tt {\tilde{\T}}

\def \S {\mathbf{S}} 
\def \T {\mathbf{T}}

\def \W {\mathbf{W}} \def \w {\mathbf{w}}

\def \z {\mathbf{z}} 
\def \det {\mathrm{det}}
\def \Tr {\mathrm{Tr}}
\def \var {\mathrm{var}}

\def \Ct {\tilde{\C}}
\def \Cr {{\bf C}}	 \def \Ct {\tilde{{\bf C}}}
\def \Crl {\Cr^{(l)}}	 \def \Ctl {\Ct^{(l)}}
\def \Crj {\Cr^{(j)}}	 \def \Ctj {\Ct^{(j)}}
\def \Crk {\Cr^{(k)}}	 \def \Ctk {\Ct^{(k)}}
\def \dr {\delta}	 \def \dt {\tilde{\delta}}
\def \drl {\delta_l}	 \def \dtl {\tilde{\delta}_l}
\def \frl {f_l}		 \def \ftl {\tilde{f}_l}

\begin{document}

\title{\Large {\bf On the Capacity Achieving Covariance Matrix for
    Frequency Selective  MIMO Channels Using the Asymptotic Approach} }

\author{Florian Dupuy, Philippe Loubaton, {\em Fellow}, {\em IEEE}\medskip\\
\today}

\maketitle


\begin{abstract}
  In this contribution, an algorithm for evaluating the capacity-achieving input covariance matrices for frequency selective Rayleigh MIMO channels is proposed.
  In contrast with the flat fading Rayleigh case, no closed-form expressions for the eigenvectors of the optimum input covariance matrix are available.
  Classically, both the eigenvectors and eigenvalues are computed numerically and the corresponding optimization algorithms remain computationally very demanding.
  
  In this paper, it is proposed to optimize (w.r.t. the input covariance matrix) a large system approximation of the average mutual information derived by Moustakas and Simon.
  The validity of this asymptotic approximation is clarified thanks to Gaussian large random matrices methods.
  It is shown that the approximation is a strictly concave function of the input covariance matrix and that the average mutual information evaluated at the argmax of the approximation is equal to the capacity of the channel up to a $\mathcal{O}\left(\frac{1}{t}\right)$ term, where $t$ is the number of transmit antennas.
  An algorithm based on an iterative waterfilling scheme is proposed to maximize the average mutual information approximation, and its convergence studied.
  Numerical simulation results show that, even for a moderate number of transmit and receive antennas, the new approach provides the same results as direct maximization approaches of the average mutual information. 
\end{abstract}


\begin{keywords}
Ergodic capacity, large random matrices, frequency selective MIMO channels
\end{keywords}

\IEEEpeerreviewmaketitle







\section{Introduction}



When the channel state information is available at both the receiver and the transmitter of a MIMO system, the problem of designing the transmitter in order to maximize the (Gaussian) mutual information of the system has been addressed successfully in a number of papers. 
This problem is however more difficult when the transmitter has the knowledge of the statistical properties of the channel, the channel state information being still available at the receiver side, a more realistic assumption in the context of mobile systems.
In this case, the mutual information is replaced by the average mutual information (EMI), which, of course, is more complicated to optimize.
\medskip

The optimization problem of the EMI has been addressed extensively in the case of certain flat fading Rayleigh channels.
In the context of the so-called Kronecker model, it has been shown by various authors (see e.g. \cite{Goldsmith-Jafar-etal-03} for a review) that the eigenvectors of the optimal input covariance matrix must coincide with the eigenvectors of the transmit correlation matrix.
It is therefore sufficient to evaluate the eigenvalues of the optimal matrix, a problem which can be solved by using standard optimization algorithms.
Similar results have been obtained for flat fading uncorrelated Rician channels (\cite{Hoesli-Kim-Lapidoth-05}).
\medskip

In this paper, we consider this EMI maximization problem in the case of popular frequency selective MIMO channels (see e.g. \cite{boelcskei}, \cite{Moustakas-Simon-07}) with independent paths.
In this context, the eigenvectors of the optimum transmit covariance matrix have no closed-form expressions, so that both the eigenvalues and the eigenvectors of the matrix have to be evaluated numerically.
For this, it is possible to adapt the approach of \cite{Vu-Paulraj-05} developed in the context of correlated Rician channels.
However, the corresponding algorithms are computationally very demanding as they heavily rely on intensive Monte-Carlo simulations.
We therefore propose to optimize the approximation of the EMI, derived by Moustakas and Simon (\cite{Moustakas-Simon-07}), in principle valid when the number of transmit and receive antennas converge to infinity at the same rate, but accurate for realistic numbers of antennas.
This will turn out to be a simpler problem.
We mention that, while \cite{Moustakas-Simon-07} contains some results related to the structure of the argument of the maximum of the EMI approximation, \cite{Moustakas-Simon-07} does not propose any optimization algorithm.
\medskip

We first review the results of \cite{Moustakas-Simon-07} related to the large system approximation of the EMI.
The analysis of \cite{Moustakas-Simon-07} is based on the so-called replica method, an ingenious trick whose mathematical relevance has not yet been established mathematically.
Using a generalization of the rigorous analysis of \cite{hachem2008anew}, we verify the validity of the approximation of \cite{Moustakas-Simon-07} and provide the convergence speed under certain technical assumptions.
Besides, the expression of the approximation depends on the solutions of a non linear system.
The existence and the uniqueness of the solutions are not addressed in \cite{Moustakas-Simon-07}.
As our optimization algorithm needs to solve this system, we clarify this crucial point.
We show in particular that the system admits a unique solution that can be evaluated numerically using the fixed point algorithm.
Next, we study the properties of the EMI approximation, and briefly justify that it is a strictly concave function of the input covariance matrix.
We show that the mutual information corresponding to the argmax of the the EMI approximation is equal to the channel capacity up to a
$\mathcal{O}\left(\frac{1}{t}\right)$ term, where $t$ is the number of transmit antennas.
Therefore it is relevant to optimize the EMI approximation to evaluate the capacity achieving covariance matrix.
We finally present our maximization algorithm of the EMI approximation.
It is based on an iterative waterfilling  algorithm which, in some sense, can be seen as a generalization of \cite{Wen-Com-06} devoted to the Rayleigh context and of \cite{dumont2010on, dumont-2007} devoted to the correlated Rician case:
Each iteration will be devoted to solve the above mentioned  system of nonlinear equations as well as a standard waterfilling problem.
It is proved that the algorithm converges towards the optimum input covariance matrix as long as it converges\footnote{Note however that we have been unable to prove formally its convergence.}. \medskip

The paper is organized as follows. Section \ref{sec:model} is devoted to the presentation of the channel model, the underlying assumptions, the problem statement.
In section \ref{sec:property}, we rigorously derive the large system approximation of the EMI with Gaussian methods and establish some properties of the asymptotic approximation as a function of the covariance matrix of the input signal.
The maximization problem of the EMI approximation is then studied in section \ref{sec:algo}.
Numerical results are provided in section \ref{sec:simulations}.

\section{Problem statement}
\label{sec:model}

\subsection{General Notations} In this paper, the notations $s$, ${\bf x}$, ${\bf M}$, stand for scalars, vectors and matrices, respectively.
As usual, $\|{\bf x} \|$ represents the Euclidian norm of vector ${\bf x}$, and $\| {\bf M} \|$, $\rho({\bf M})$ and $| {\bf M} |$ respectively stand for the spectral norm, the spectral radius and the determinant of matrix ${\bf M}$.
The superscripts $(.)^T$ and $(.)^H$ represent respectively the transpose and transpose conjugate.
The trace of ${\bf M}$ is denoted by $\mathrm{Tr}({\bf M})$.
The mathematical expectation operator is denoted by $\mathbb{E}(\cdot)$.
We denote by $\delta_{i,j}$ the Kronecker delta, i.e. $\delta_{i,j}=1$ if $i=j$ and $0$ otherwise.

All along this paper, $r$ and $t$ stand for the number of receive and transmit antennas.
Certain quantities will be studied in the asymptotic regime $t \to \infty$, $r \to \infty$ in such a way that $t/r \to c \in (0,\infty)$.
In order to simplify the notations, $t \to \infty$ should be understood from now on as $t \to \infty$, $r \to \infty$ and $t/r \to c \in (0,\infty)$. A matrix ${\bf M}_t$ whose size depends on $t$ is said to be uniformly bounded if $\sup_{t} \| {\bf M}_t \| < \infty$.

Several variables used throughout this paper depend on various parameters, e.g. the number of antennas, the noise level, the covariance matrix of the transmitter, etc.
In order to simplify the notations, we may not always mention all these dependencies.


\subsection{Channel model}

We consider a wireless MIMO link with $t$ transmit and $r$ receive
antennas corrupted by a multi-paths propagation channel. The discrete-time propagation channel between the transmitter and the receiver 
is characterized by the input-output equation
\begin{equation}
\label{eq:model-signaux}
{\bf y}(n) = \sum_{l=1}^{L} \Hl {\bf s}(n-l+1) + {\bf n}(n)= [{\bf H}(z)]{\bf s}(n) + {\bf n}(n),
\end{equation}
where ${\bf s}(n) = s_1(n), \ldots, s_t(n))^{T}$ and ${\bf y}(n) = (y_1(n), \ldots, y_r(n))^{T}$ 
represent the transmit and the receive vector at time $n$ respectively. ${\bf n}(n)$ is
an additive Gaussian noise such that $\mathbb{E}({\bf n}(n){\bf n}(n)^{H}) = \sigma^{2} {\bf I}$.  ${\bf H}(z)$ denotes
the transfer function of the discrete-time equivalent channel defined by
\begin{equation}
\label{eq:def-transfert}
{\bf H}(z) = \sum_{l=1}^{L} \Hl \, z^{-(l-1)}.
\end{equation}
Each coefficient $\Hl$ is assumed to be a Gaussian random matrix given by
\begin{equation}
\label{eq:structure-Hl}
\Hl = \frac{1}{\sqrt{t}} (\Crl)^{1/2} {\bf W}_l (\Ctl)^{1/2},
\end{equation}
where ${\bf W}_l$ is a $r \times t$ random matrix whose entries are independent and identically distributed complex circular Gaussian random variables, with zero mean and unit variance.
The matrices $\Crl$ and $\Ctl$ are positive definite, and respectively account for the receive and transmit antenna correlation.
This correlation structure is called a separable or Kronecker correlation model.
We also assume that for each $k \neq l$, matrices $\Hk$ and $\Hl$ are independent.
Note that our assumptions imply that $\Hl\neq0$ for $l=1,\ldots,L$. However, it can be checked easily that the results stated in this paper remain valid if some coefficients $(\Hl)_{l=1,\ldots,L}$ are zero.

In the context of this paper, the channel matrices are assumed perfectly known at the receiver side.
However, only the statistics of the $(\Hl)_{l=1, \ldots, L}$, i.e. matrices $(\Ctl, \Crl)_{l=1, \ldots, L}$, are available at the transmitter side.

\subsection{Ergodic capacity of the channel.}
Let ${\bf Q}(e^{2 i \pi \nu})$ be the $t \times t$ spectral density matrix of the transmit signal ${\bf s}(n)$, which is assumed to verify the transmit power condition
\begin{equation}
\label{eq:power}
\frac{1}{t} \int_0^1 \mathrm{Tr}({\bf Q}(e^{2 i \pi \nu})) d \nu = 1.
\end{equation}
Then, the (Gaussian) ergodic mutual information $I({\bf Q}(.))$ between the transmitter and the receiver is defined as
\begin{equation}
\label{eq:def-EMI}
I({\bf Q}(.)) = \mathbb{E}_{\cal{W}} \left[ \int_0^1 \log \left| {\bf I}_r + \frac{1}{\sigma^{2}} {\bf H}(e^{2 i \pi \nu}) {\bf Q}(e^{2 i \pi \nu}) {\bf H}(e^{2 i \pi \nu})^{H} \right| \, d\nu \right],
\end{equation}
where $\mathbb{E}_{\cal{W}}[.]=\mathbb{E}_{({\bf W}_l)_{l=1, \ldots, L}}[.]$. The ergodic capacity of the MIMO channel is equal to the maximum of $I({\bf Q}(.))$ over the set of all spectral density matrices 
satisfying the constraint (\ref{eq:power}). The hypotheses formulated on the statistics of the channel allow however to 
limit the optimization to the set of positive matrices which are independent of the frequency $\nu$. This is because the 
probability distribution of matrix ${\bf H}(e^{2 i \pi \nu})$ is clearly independent of the frequency $\nu$.
More precisely, the mutual information ${\bf I}({\bf Q}(.))$ is also given by
\[
I({\bf Q}(.)) = \mathbb{E}_{\bf H} \left[ \int_0^1 \log \left| {\bf I}_r + \frac{1}{\sigma^{2}} {\bf H}(1) {\bf Q}(e^{2 i \pi \nu}) {\bf H}(1)^{H} \right| \, d\nu \right],
\]
where ${\bf H} = \sum_{l=1}^{L} \Hl={\bf H}(1)$. Using the concavity of the logarithm, we obtain that
\[
I({\bf Q}(.))  \leq  \mathbb{E}_{\bf H} \left[ \log \left| {\bf I}_r + \frac{1}{\sigma^{2}} {\bf H}(1) \left( \int_0^1 {\bf Q}(e^{2 i \pi \nu}) d\nu \right) \, {\bf H}(1)^{H} \right| \,  \right].
\]
We denote by ${\cal C}$ the cone of non negative hermitian matrices, and by ${\cal C}_1$ the subset of all matrices ${\bf Q}$ of ${\cal C}$ satisfying 
$\frac{1}{t} \mathrm{Tr}({\bf Q}) = 1$. If ${\bf Q}$ is an element of ${\cal C}_1$, the mutual information $I({\bf Q})$ reduces to  
\begin{equation}
I({\bf Q}) =  \mathbb{E}_{\bf H} \left[\log \left| {\bf I}_r + \frac{1}{\sigma^{2}} {\bf H} {\bf Q}  {\bf H}^{H} \right| \right].
\label{eq:expre-finale-I} 
\end{equation}
${\bf Q} \mapsto I({\bf Q})$ is strictly concave on the convex set ${\cal C}_1$ and reaches its maximum at a unique element ${\bf Q}_* \in {\cal C}_1$. It is clear that 
if ${\bf Q}(e^{2 i \pi \nu})$ is any spectral density satisfying (\ref{eq:power}), then the matrix $ \int_0^1 {\bf Q}(e^{2 i \pi \nu}) d\nu $ 
is an element of ${\cal C}_1$. Therefore, 
\[
\mathbb{E}_{\bf H}  \left[ \log \left| {\bf I}_r + \frac{1}{\sigma^{2}} {\bf H} \left( \int_{0}^{1} {\bf Q}(e^{2 i \pi \nu}) d\nu \right) \, {\bf H}^{H} \right| \,  \right] \leq  \mathbb{E}_{\bf H} \left[\log \left| {\bf I}_r + \frac{1}{\sigma^{2}} {\bf H} {\bf Q}_{*}   {\bf H}^{H} \right| \right].
\]
In other words, 
\[
I({\bf Q}(.)) \leq I({\bf Q}_{*})
\]
for each spectral density matrix verifying  (\ref{eq:power}). This shows that the maximum of function $I$ over the set of all spectral densities satisfying (\ref{eq:power}) is reached on the set ${\cal C}_1$. The 
ergodic capacity ${\cal C}_E$ of the channel is thus equal to 
\begin{equation}
  {\cal C}_E  = \max_{{\bf Q} \in {\cal C}_1} I({\bf Q}).
  \label{eq:max_C1}
\end{equation}
We note that property \eqref{eq:max_C1} also holds if the time delays of the channel are non integer multiples of the symbol period, provided that the receiving filter coincides with the ideal low-pass filter on the $[-\frac{1}{2T}, \frac{1}{2T}]$ frequency interval, where $T$ denotes the symbol period.
If this is the case, the transfer function $\H(e^{2 i \pi \nu})$ is equal to $\H(e^{2 i \pi \nu})=\sum_{l=1}^L \Hl e^{-2 i \pi \nu \tau_l }$, where 
$\tau_l$ is the delay associated to path $l$ for $l=1, \ldots, L$.
The probability distribution of $\H(e^{2 i \pi \nu})$ does not depend on $\nu$ and this leads immediately to \eqref{eq:max_C1}.
If the matrices $(\Crl)_{l=1, \ldots, L}$ all coincide with a matrix ${\bf C}$, matrix ${\bf H}$ follows a Kronecker model with transmit and receive covariance matrices $\frac{1}{L} \sum_{l=1}^{L} \Ctl$ and ${\bf C}$ respectively \cite{Cedric-GC}.
In this case, the eigenvectors of the optimum matrix ${\bf Q}_*$ coincide with the eigenvectors of $\frac{1}{L}\sum_{l=1}^{L} \Ctl$.
The situation is similar if the transmit covariance matrices $(\Ctl)_{l=1, \ldots, L}$ coincide.
In the most general case, the eigenvectors of ${\bf Q}_*$ have however no closed-form expression. The evaluation of ${\bf Q}_*$ and of the channel capacity ${\cal C}_E$ is thus a more difficult problem.
A possible solution consists in adapting the Vu-Paulraj approach (\cite{Vu-Paulraj-05}) to the present context.
However, the algorithm presented in \cite{Vu-Paulraj-05} is very demanding since the evaluations of the gradient and the Hessian of $I({\bf Q})$ require intensive Monte-Carlo simulations.

\subsection{The large system approximation of $I({\bf Q})$}
When $t$ and $r$ converge to $\infty$ while $t/r \rightarrow c$, $c \in (0, \infty)$, 
\cite{Moustakas-Simon-07} showed that $I({\bf Q})$ can be approximated by $\overline{I}({\bf Q})$ defined by
\begin{equation}
\label{eq:expre-Ibarre}
  \overline{I}({\bf Q}) = \log \left| {\bf I} + \sum_{l=1}^{L} \dtl({\bf Q}) \Crl \right|
   +\log \left|{\bf I} + {\bf Q} \left( \sum_{l=1}^{L} \drl({\bf Q}) \Ctl \right) \right|
   -\sigma^{2} t \left( \sum_{l=1}^{L}  \drl({\bf Q})  \dtl({\bf Q}) \right)
\end{equation} 
where $(\dr_1({\bf Q}),\ldots, \dr_L({\bf Q}))^T={\bs \dr({\bf Q})}$ and $(\dt_1({\bf Q}), \ldots,\dt_L({\bf Q}))^T={\bs \dt({\bf Q})}$ are the positive solutions of the system of $2L$ equations:
\begin{equation}
\left\{\begin{array}{l}
\kappa_l = \frl(\tilde{\bs{\kappa}})
\vspace{1mm}\\
\tilde{\kappa}_l = \ftl(\bs{\kappa}, {\bf Q})
\end{array}\right. \text{ for } l=1, \ldots, L,
\label{eq:canonique}
\end{equation}
with ${\bs \kappa} = (\kappa_1, \ldots, \kappa_L)^{T}$ and  $\tilde{{\bs \kappa}} = (\tilde{\kappa}_1, \ldots, \tilde{\kappa}_L)^{T}$, and with
\begin{equation}
\label{eq:canoniquebis}
\left\{\begin{array}{l}
  \frl(\tilde{\bs{\kappa}}) = \frac{1}{t} \mathrm{Tr} \left[ \Crl {\bf T}(\tilde{\bs{\kappa}}) \right],
  \vspace{1mm}\\
  \ftl(\bs{\kappa}, {\bf Q}) = \frac{1}{t} \mathrm{Tr} \left[ {\bf Q}^{1/2} \Ctl {\bf Q}^{1/2}  \tilde{\bf T}(\bs{\kappa}, {\bf Q}) \right],
\end{array} \right.
\end{equation}
where $r \times r$ matrix $\T(\tilde{\bs\kappa})$ and $t \times t$ matrix $\tilde\T(\bs\kappa,\Q)$ are defined by
\begin{equation}
  \label{eq:canoniqueter}
  \left\{\begin{array}{l}
    {\bf T}(\tilde{\bs{\kappa}})=\left[\sigma^{2} \left({\bf I} + \sum_{j=1}^{L} \tilde{\kappa}_j \Crj \right)\right]^{-1},
    \vspace{1mm}\\
    \tilde{\bf T}(\bs{\kappa}, {\bf Q})=\left[\sigma^{2} \left( {\bf I} + \sum_{j=1}^{L} \kappa_j  {\bf Q}^{1/2} \Ctj  {\bf Q}^{1/2} \right)\right]^{-1}.
  \end{array} \right.
\end{equation}

\section{Deriving the large system approximation}
\label{sec:property}

\subsection{The canonical equations}
\label{sec:canon-eq}
In \cite{Moustakas-Simon-07}, the existence and the uniqueness of positive solutions to (\ref{eq:canonique}) is assumed without justification.
Moreover no algorithm is given for the calculation of the $\drl$ and $\dtl$, $l=1,\ldots, L$.
We therefore clarify below these important points.
We consider the case ${\bf Q} = {\bf I}$ in order to simplify the notations.
To address the general case it is sufficient to change matrices $(\Ctl)_{l=1, \ldots, L}$ into  $({\bf Q}^{1/2}\Ctl {\bf Q}^{1/2})_{l=1, \ldots, L}$ in what follows.

\begin{theorem}
  \label{th:ex-un-algo-cano}
  The system of equations (\ref{eq:canonique}) admits unique positive solutions $(\drl)_{l=1, \ldots, L}$ and $(\dtl)_{l=1, \ldots, L}$,
  which are the limits of the following fixed point algorithm:
	\begin{itemize}
	\item[-] Initialization: $\delta_l^{(0)} > 0$,  $\tilde{\delta}_l^{(0)} > 0$, $l=1,\ldots, L$.
	\item[-] Evaluation of the $\drl^{(n+1)}$ and $\dtl^{(n+1)}$ from ${\bs{\delta}}^{(n)}=(\delta_1^{(n)}, \ldots, \delta_L^{(n)})^T$ and $\tilde{\bs{\delta}}^{(n)}=(\tilde{\delta}_1^{(n)}, \ldots, \tilde{\delta}_L^{(n)})^{T}$:
	\begin{eqnarray}
	\label{eq:point-fixe}
	\left\{\begin{array}{l}
	  \delta_l^{(n+1)} = \frl(\tilde{\bs{\delta}}^{(n)}),
	  \vspace{1mm}\\
	  \tilde{\delta}_l^{(n+1)} = \ftl(\bs{\delta}^{(n)}, {\bf I}).
	\end{array}\right.
	\end{eqnarray}
	\end{itemize}
\end{theorem} \medskip

\begin{proof}
We prove the existence and uniqueness of positive solutions.
\begin{enumerate}
 \item{\it Existence}:
	Using analytic continuation technique, we show in Appendix \ref{apx:delta_exist-unique} that the fixed point algorithm introduced converges to positive coefficients $\drl$ and $\dtl$, $l=1,\ldots,L$.
	As functions $\bs{\tilde\kappa} \mapsto f_l(\bs{\tilde\kappa})$ and $\bs{\kappa} \mapsto \tilde f_l(\bs{\kappa},{\bf I})$ are clearly continuous, the limit of $({\bs{\delta}}^{(n)},\tilde{\bs{\delta}}^{(n)})$ when $n\rightarrow\infty$
	satisfies equation (\ref{eq:canonique}).
	Hence, the convergence of the algorithm yields the existence of a positive solution to (\ref{eq:canonique}).\medskip
	
 \item{\it Uniqueness}:
	Let $(\bs{\dr},\bs{\dt})$ and $(\bs{\dr}',\bs{\dt}')$ be two solutions of the canonical equation (\ref{eq:canonique}) with ${\bf Q}={\bf I}$. We denote $({\bf T}, \tilde{\bf T})$ and $({\bf T}', \tilde{\bf T}')$ the associated matrices defined by (\ref{eq:canoniqueter}), where $(\bs\kappa,\bs{\tilde\kappa})$ respectively coincide with $(\bs\delta,\bs{\tilde\delta})$ and $(\bs\delta',\bs{\tilde\delta}')$.
	Introducing ${\bf e}=\bs{\dr}-\bs{\dr}'=(e_1,\ldots,e_L)^T$ we have:
	\begin{align}
	  e_l
	  &= \frac{1}{t}\mathrm{Tr}\left[\Crl{\bf T}({\bf T'}^{-1}-{\bf T}^{-1}){\bf T'}\right] \notag
	  \\
	  &= \frac{\sigma^2}{t}\sum_{k=1}^L(\tilde{\delta}_k'-\tilde{\delta}_k)\mathrm{Tr}\left(\Crl{\bf T}\Crk{\bf T'}\right).
	  \label{el:eqn}
	\end{align}
	Similarly, with $\tilde{\bs{e}}=\bs{\dt}-\bs{\dt}'=(\tilde{e}_1,\ldots,\tilde{e}_L)^T$,
	\begin{equation}
	  \tilde{e}_k = \frac{\sigma^2}{t}\sum_{l=1}^L(\delta_l'-\delta_l)\mathrm{Tr}\left(\Ctk\tilde{\bf T}\Ctl\tilde{\bf T'}\right).
	  \label{elt:eqn}
	\end{equation}
	And (\ref{el:eqn}) and (\ref{elt:eqn}) can be written together as
	\begin{equation}
	  \left[\begin{array}{cc} {\bf I} & \sigma^2{\bf A}({\bf T},{\bf T'}) \\ \sigma^2\tilde{\bf A}(\tilde{\bf T},\tilde{\bf T'}) & {\bf I}  \end{array}\right] 
	  \left[\begin{array}{c}{\bf e} \\ \tilde{\bf e} \end{array}\right]
	  = {\bf 0},
	  \label{unq:matrix}
	\end{equation}
	with ${\bf A}_{kl}({\bf T},{\bf T'})=\frac{1}{t}\mathrm{Tr}\left(\Crk{\bf T}\Crl{\bf T'}\right)$ and $\tilde{\bf A}_{kl}(\tilde{\bf T},\tilde{\bf T'})=\frac{1}{t}\mathrm{Tr}(\Ctk\tilde{\bf T}\Ctl\tilde{\bf T'})$.
	We will now prove that $\rho({\bf M})<1$, with ${\bf M}=\sigma^4\tilde{\bf A}(\tilde{\bf T},\tilde{\bf T}'){\bf A}({\bf T},{\bf T'})$. This will imply that the matrix governing the linear system \eqref{unq:matrix} is invertible, and thus that ${\bf e}=\tilde{\bf e}={\bf 0}$, i.e. the uniqueness.
	\begin{align}
	  |{\bf M}_{kl}| &= \left|\frac{\sigma^4}{t^2}\sum_{j=1}^L\mathrm{Tr}(\Ctk\tilde{\bf T}\Ctj\tilde{\bf T'})\mathrm{Tr}(\Crj{\bf T}\Crl{\bf T'})\right| \notag
	  \\ &\leq \frac{\sigma^4}{t^2}\sum_{j=1}^L\left|\mathrm{Tr}(\Ctk\tilde{\bf T}\Ctj\tilde{\bf T'})\right|\left|\mathrm{Tr}(\Crj{\bf T}\Crl{\bf T'})\right|.
	  \label{Mkl-ineq}
	\end{align}
	Thanks to the inequality $|\mathrm{Tr}({\bf AB})|\leq\sqrt{\mathrm{Tr}({\bf AA^H})\mathrm{Tr}({\bf BB^H})}$, we have
	\begin{align}
	  \left\{\begin{array}{l}
	    \displaystyle
	    \frac{1}{t}\left|\mathrm{Tr}(\Ctk\tilde{\bf T}\Ctj\tilde{\bf T'})\right| \leq 
	    \sqrt{\tilde{\bf A}_{kj}(\tilde{\bf T},\tilde{\bf T})\tilde{\bf A}_{kj}(\tilde{\bf T}',\tilde{\bf T'})},  
	    \vspace{3mm}\\
	    \displaystyle
	    \frac{1}{t}\left|\mathrm{Tr}(\Crj{\bf T}\Crl{\bf T'})\right|
	    \leq \sqrt{{\bf A}_{jl}({\bf T},{\bf T}){\bf A}_{jl}({\bf T}',{\bf T'})}.
	  \end{array}\right.
	  \label{eq:Tr_CTCT}
	\end{align}
	Using \eqref{eq:Tr_CTCT} in \eqref{Mkl-ineq} gives
	\[
	  |{\bf M}_{kl}| \leq \sigma^4\sum_{j=1}^L\sqrt{\tilde{\bf A}_{kj}(\tilde{\bf T})\tilde{\bf A}_{kj}(\tilde{\bf T'}){\bf A}_{jl}({\bf T}){\bf A}_{jl}({\bf T}')},
	\]
	where matrices $\A(\T)$ and $\tilde\A(\tilde\T)$ are defined by
	\begin{equation}
	  \label{eq:A-Atilde}
	  \left\{\begin{array}{l}
	    \displaystyle {\bf A}_{kl}({\bf T})=\frac{1}{t}\mathrm{Tr} (\Crk{\bf T}\Crl{\bf T}) ={\bf A}_{kl}({\bf T},{\bf T}),
	    \vspace{3mm}\\
	    \displaystyle \tilde{\bf A}_{kl}(\tilde{\bf T})=\frac{1}{t}\mathrm{Tr}(\Ctk\tilde{\bf T}\Ctl\tilde{\bf T})=\tilde{\bf A}_{kl}(\tilde{\bf T},\tilde{\bf T}).
	  \end{array}\right.
	\end{equation}
	Using Cauchy-Schwarz inequality,
	\begin{align}
	  |{\bf M}_{kl}| &\leq \sigma^4\sqrt{\bigg(\sum_{j=1}^L\tilde{\bf A}_{kj}(\tilde{\bf T}){\bf A}_{jl}({\bf T})\bigg)\bigg(\sum_{j=1}^L\tilde{\bf A}_{kj}(\tilde{\bf T'}){\bf A}_{jl}({\bf T'})\bigg)}. \notag
	\end{align}
	Hence, defining the matrix $\bf{P}$ by
	${\bf P}_{kl}=\sqrt{(\sigma^4\tilde{\bf A}(\tilde{\bf T}){\bf A}({\bf T}))_{kl}}\sqrt{(\sigma^4\tilde{\bf A}(\tilde{\bf T'}){\bf A}({\bf T}'))_{kl}}$, we have $|{\bf M}_{kl}|\leq{\bf P}_{kl}$ $\forall k,l$.
	Theorem 8.1.18 of \cite{horn1990matrix} then yields $\rho({\bf M})\leq\rho({\bf P})$.
	Besides, Lemma 5.7.9 of \cite{horn1994topics} used on the definition of ${\bf P}$ gives:
	\begin{equation}
	 \rho({\bf P})\leq\sqrt{\rho\left(\sigma^4\tilde{\bf A}(\tilde{\bf T}){\bf A}({\bf T})\right)}\sqrt{\rho\left(\sigma^4\tilde{\bf A}(\tilde{\bf T'}){\bf A}({\bf T}')\right)}.
	 \label{eq:rsp-hada}
	\end{equation}
	Lemma \ref{lm:rsp} \eqref{it:rspAAt} in Appendix \ref{apx:alpha-delta} implies that
	$\rho(\sigma^4\tilde{\bf A}(\tilde{\bf T}){\bf A}({\bf T}))<1$ and $\rho(\sigma^4\tilde{\bf A}(\tilde{\bf T'}){\bf A}({\bf T'}))<1$,
	so that (\ref{eq:rsp-hada}) finally implies:
	\[\rho({\bf M})\leq\rho({\bf P})<1.\]
\end{enumerate}
This completes the proof of Theorem \ref{th:ex-un-algo-cano}.
\end{proof}\medskip

\subsection{Deriving the approximation of $I(\Q=\I_t)$ with Gaussian methods}
\label{sec:appGauss}
We consider in this section the case $\Q = \I_t$. We note $I=I(\I_t)$, $\overline{I}=\overline{I}(\I_t)$.
We have proved in the previous section the consistency of $\overline{I}(\Q)$ definition.
To establish the approximation of $I(\Q)$, \cite{Moustakas-Simon-07} used the replica method, a useful and simple trick whose mathematical relevance is not yet proved in the present context.
Moreover, no assumptions were specified for the convergence of $I(\Q)$ towards $\overline{I}(\Q)$.
However, using large random matrix techniques similar to those of \cite{hachem2008anew}, \cite{dumont2010on}, it is possible to prove rigorously the following theorem, in which the (mild) suitable technical assumptions are clarified.
\medskip

\begin{theorem}
\label{th:appr}
 Assume that, for every $j \in \left\{ 1, \ldots, L\right\}$, $\sup_t\|\Crj\|<+\infty$, $\sup_t\|\Ctj\|<+\infty$,
 $\inf_t\left(\frac{1}{t}\mathrm{Tr\,}\Crj\right)>0$ and $\inf_t\left(\frac{1}{t}\mathrm{Tr}\,\Ctj\right)>0$.
 Then,
  \[ I=\overline{I}+\mathcal{O}\left(\frac{1}{t}\right).\]
\end{theorem} \medskip
\begin{sktchproof}
 The proof is done in three steps:
 \begin{enumerate}
  \item 
	In a first step we derive a large system approximation of $\mathbb{E}_\H[\mathrm{Tr}\,{\bf S}]$, where ${\bf S}=\left({\bf H}{\bf H}^H+\sigma^2{\bf I}_r\right)^{-1}$ is the resolvent of $\H\H^H$ at point $-\sigma^2$.
	Nonetheless the approximation is expressed with the terms
	$\alpha_l= \frac{1}{t} \mathbb{E}_\H\left[ \mathrm{Tr} \left( \Crl {\bf S} \right) \right]$, $l=1,\ldots,L$,
	which still depend on the entries of $\mathbb{E}_\H[ \S ]$.
  \item
	A second step refines the previous approximation to obtain an approximation which this time only depends on the variance structure of the channels, i.e. matrices $(\Crl)_{\ell \in \{1,\ldots, L\}}$ and $(\Ctl)_{\ell \in \{1,\ldots, L\}}$.
  \item
	The previous approximation is used to get the asymptotic behavior of mutual information by a proper integration.
 \end{enumerate}
\end{sktchproof}

\begin{proof} We now sketch the three steps stated above. We provide the missing details in the Appendix.
\subsubsection{A first large system approximation of $\mathbb{E}_\H[\mathrm{Tr}\,{\bf S}]$}
 We introduce vectors ${\bs \alpha} = (\alpha_1, \ldots, \alpha_L)^{T}$ and $\tilde{\bs \alpha} = (\tilde\alpha_1, \ldots, \tilde\alpha_L)^{T}$ defined by
 \begin{equation}
 \left\{\begin{array}{l}
   \alpha_l= \frac{1}{t} \mathrm{Tr} \left[ \Crl \mathbb{E}_\H[{\bf S}] \right]
   \vspace{1mm}\\
   \tilde\alpha_l= \frac{1}{t} \mathrm{Tr} \left[ \Ctl \tilde\R \right]
  \end{array} \right. \mathrm{for} \ l=1,\ldots,L,
  \label{eq:def-alphas}
 \end{equation}
 where matrix $\tilde\R$ is defined by 
 $
  \tilde\R(\bs{\alpha})=\left[\sigma^{2} \left( {\bf I} + \sum_{j=1}^{L} \alpha_j  \Ctj  \right)\right]^{-1}
 $.
 Using large random matrix techniques similar to those of \cite{hachem2008anew,dumont2010on}, the following proposition is proved in Appendix \ref{apx:1st_apprx}.
 \begin{proposition}
 \label{prop:S-appr}
 Assume that, for every $j \in \left\{ 1, \ldots, L\right\}$, $\sup_t\|\Crj\|<+\infty$, $\sup_t\|\Ctj\|<+\infty$.
 Then $\mathbb{E}_\H[ \S ]$ can be written as
  \begin{equation}
   \mathbb{E}_\H[ \S] ={\bf R} + \bs \Upsilon,
   \label{eq:propS-appr}
  \end{equation}
  where matrix $\bs \Upsilon$ is such that
  $\frac{1}{t}\mathrm{Tr}(\bs \Upsilon \A)= \mathcal{O}\left(\frac{1}{t^2}\right)$ for any uniformly bounded matrix ${\bf A}$
  and matrix ${\bf R}$ is defined by
  ${\bf R}(\bs{\tilde\alpha})=\left[\sigma^{2} \left( {\bf I} + \sum_{j=1}^{L} \tilde\alpha_j  \Crj  \right)\right]^{-1}$.
 \end{proposition}
 One can check that the entries of matrix $\bs\Upsilon$ are $\mathcal{O}\left(\frac{1}{t^{3/2}}\right)$; nevertheless this result is not needed here.
 It follows from Proposition~\ref{prop:S-appr} that, for any $r \times r$ matrix ${\bf A}$ uniformly bounded in $r$,
 \begin{equation}
  \frac{1}{t}\mathbb{E}_\H[\mathrm{Tr}({\bf S A})] =\frac{1}{t}\mathrm{Tr}({\bf R A}) + \mathcal{O}\left(\frac{1}{t^2}\right).
  \label{eq:trSA}
 \end{equation}
 Taking ${\bf A}={\bf I}$ gives a first approximation of $\mathbb{E}_\H[\mathrm{Tr}\,{\bf S}]$:
 \begin{equation}
  \mathbb{E}_\H[\mathrm{Tr}\,{\bf S}] =\mathrm{Tr}\,{\bf R } + \mathcal{O}\left(\frac{1}{t}\right).
  \label{eq:1st_appr}
 \end{equation}
 Nonetheless matrix ${\bf R}$ depends on $\mathbb{E}_\H[{\bf S}]$ through vector $\bs{\alpha}$.
 
 \subsubsection{A refined large system approximation of $\mathbb{E}_\H[\mathrm{Tr}\ {\bf S}]$}
 We first recall from paragraph \ref{sec:canon-eq} that ${\bf T}$ is the matrix defined by (\ref{eq:canoniqueter}) associated to the solutions $(\bs{\delta},\bs{\tilde\delta})$ of the canonical equation (\ref{eq:canonique}) with ${\bf Q}={\bf I}$:
 $\T~=~\left(\sigma^2\left(\I_r+\sum_{l=1}^L \tilde\delta_l \Crl \right)\right)^{-1}$.
 We introduce the following proposition which will lead to the desired approximation of $\mathbb{E}_\H[\mathrm{Tr}\ {\bf S}]$:
 \begin{proposition}
 \label{prop:2nd-appr}
 Assume that, for every $j \in \left\{ 1, \ldots, L\right\}$, $\sup_t\|\Crj\|<+\infty$, $\sup_t\|\Ctj\|<+\infty$,
 $\inf_t\left(\frac{1}{t}\mathrm{Tr\,}\Crj\right)>0$ and $\inf_t\left(\frac{1}{t}\mathrm{Tr}\,\Ctj\right)>0$.
 Let ${\bf A}$ be a $r \times r$ matrix uniformly bounded in $r$, then 
  \begin{equation}
   \frac{1}{t}\mathrm{Tr}({\bf R A})=\frac{1}{t}\mathrm{Tr}({\bf T A}) + \mathcal{O}\left(\frac{1}{t^2}\right).
   \label{eq:prop2nd-appr}
  \end{equation}
 \end{proposition}\medskip
 The proof is given in Appendix \ref{apx:alpha-delta}.
 It relies on the similarity of the systems of equations verified by the $(\alpha_l,\tilde\alpha_l)$ and the $(\delta_l,\tilde\delta_l)$.
 Actually, taking ${\bf A}=\Crl$ in (\ref{eq:trSA}) yields
 $\alpha_l= \frac{1}{t}\mathrm{Tr}(\Crl {\bf R}) + \mathcal{O}\left(\frac{1}{t^2}\right)$ and therefore
 \begin{equation}
 \left\{\begin{array}{l}
   \alpha_l= \frac{1}{t}\mathrm{Tr}\left[\Crl \left[\sigma^{2} \left( {\bf I} + \sum_{j=1}^{L} \tilde\alpha_j  \Crj  \right)\right]^{-1}\right] + \mathcal{O}\left(\frac{1}{t^2}\right)
   \vspace{1mm}\\
   \tilde\alpha_l= \frac{1}{t} \mathrm{Tr} \left[ \Ctl \left[\sigma^{2} \left( {\bf I} + \sum_{j=1}^{L} \alpha_j  \Ctj  \right)\right]^{-1} \right]
  \end{array} \right.\mathrm{for}\ l=1,\ldots,L.
 \end{equation}
 Taking $\A=\I_r$ in \eqref{eq:prop2nd-appr} together with \eqref{eq:1st_appr} leads to 
 \begin{equation}
  \mathbb{E}_\H[\mathrm{Tr}\,{\bf S}] =\mathrm{Tr}\,\T + \mathcal{O}\left(\frac{1}{t}\right)
  \label{eq:2nd_appr}
 \end{equation}

 \subsubsection{The resulting large system approximation of $I$}
 The ergodic mutual information $I$ can be written in terms of the resolvent $\S$:
 \[
  I =\mathbb{E}_\H \left[ \log \left| \I_r + \frac{\H\H^H}{\sigma^2}\right| \right] = \mathbb{E}_\H \left[ \log \left| \sigma^2 \S(\sigma^2) \right|^{-1} \right].
 \]
 As the differential of $g(\A)=\log\left| \A \right|$ is given by $g(\A+\mathbf{\delta A})=g(\A)+\mathrm{Tr}[\A^{-1}\mathbf{\delta A}]+o(\|\mathbf{\delta A}\|)$, we have the following equality:
 \[
   \frac{d I}{d\sigma^2}=-\mathbb{E}_\H\left[\frac{\mathrm{Tr}[\S(\sigma^2)\H\H^H]}{\sigma^2}\right]=-\mathbb{E}_\H\left[\frac{\mathrm{Tr}[\I_r-\sigma^2\S(\sigma^2)]}{\sigma^2}\right],
 \]
 where the last equality follows from the so-called resolvent identity 
 \begin{equation}
  \sigma^2\S(\sigma^2)=\I_r-\S(\sigma^2)\H\H^H.
  \label{eq:res-id}
 \end{equation}
  The resolvent identity is inferred easily from the definition of $\S(\sigma^2)$.
  As $I(\sigma^2=+\infty) = 0$, we now have the following expression of mutual information:
  \[
    I(\sigma^2) = \int_{\sigma^2}^{+\infty} \left( \frac{r}{\rho} - \mathbb{E}_\H \left[ \mathrm{Tr}\ \S(\rho) \right] \right) d\rho.
  \]
  This equality clearly justifies the search of a large system equivalent of $\mathbb{E}_\H \left[\mathrm{Tr}\,\S\right]$ done in the previous sections.
  Using \eqref{eq:2nd_appr}, the term under the integral sign becomes:
  \[
   \frac{r}{\sigma^2} - \mathbb{E}_\H \left[ \mathrm{Tr}\,\S \right]
   = t\sum_{l=1}^L \dtl \drl
    + \mathbb{E}_\H \left[ \mathrm{Tr} \left(\T - \S \right) \right],
    \label{eq:trm-integ-I}
  \]
  as
  $\frac{r}{\sigma^2} - \mathrm{Tr}\,\T 
  =\mathrm{Tr}\left[\left( (\sigma^2\T)^{-1} -\I_r \right)\T\right]
  =\mathrm{Tr}\left[\left(\sum_l \dtl \Crl \right)\T\right]
  =t\sum_l \dtl \drl$.
  We need to integrate $\varepsilon(t,\sigma^2)=\mathbb{E}_\H \left[ \mathrm{Tr} \left(\T - \S \right) \right]$ on $(\rho>0,+\infty)$ with respect to $\sigma^2$.
  We therefore introduce the following proposition:\medskip
  \begin{proposition}
    \label{prop:integrable}
    $\varepsilon(t,\sigma^2)=\mathbb{E}_\H \left[ \mathrm{Tr}\left( \T - \S \right) \right]$ is integrable with respect to  $\sigma^2$ on $\left(\rho>0,+\infty\right)$ and
    \[
      \int_{\rho}^{+\infty} \varepsilon(t,\sigma^2) d\sigma^2 =\mathcal{O}\left( \frac{1}{t} \right)    
    \]
  \end{proposition}\medskip
  \begin{proof}
    We prove in Appendix \ref{apx:integrable} that there exists $t_0$ 
    such that, for $t > t_0$,
    $\left| \varepsilon(t,\sigma^2)\right| \leq \frac{1}{\sigma^8 t} P\left( \frac{1}{\sigma^2} \right)$,
    where $P$ is a polynomial whose coefficients are real positive and do not depend on $\sigma^2$ nor on $t$.
    Therefore
    $\int_{\rho}^{+\infty} \varepsilon(t,\sigma^2)d\sigma^2=\mathcal{O}\left( \frac{1}{t} \right)$.
    
  \end{proof}
  We now prove that the term $t\sum_l \dtl \drl$ corresponds to the derivative of $\overline{I}(\sigma^2)$ with respect to $\sigma^2$.
  To this end, we consider the function $\mathcal{V}_0(\sigma^2,{\bs \kappa},\tilde{\bs \kappa})$ defined by
  \begin{equation}
   \mathcal{V}_0(\sigma^2,{\bs \kappa},\tilde{\bs \kappa}) =
   \log | {\bf I} + \C(\tilde{\bs \kappa}) |
   +\log |{\bf I} + \tilde\C({\bs \kappa}) |
   -\sigma^{2} t \sum_{l=1}^{L}  \kappa_l \tilde\kappa_l,
   \label{def:V0}
  \end{equation}
  where
  $\tilde{\bf C}({\bs \kappa})=\sum_{l=1}^{L} \kappa_l \Ctl$ and ${\bf C}(\tilde{\bs \kappa})=\sum_{l=1}^{L} \tilde\kappa_l \Crl$. 
  Note that $\mathcal{V}_0(\sigma^2,{\bs \dr},{\bs \dt})=\overline{I}(\sigma^2)$.
  The derivative of $\overline{I}(\sigma^2)$ can then be expressed in terms of the partial derivatives of ${\cal V}_0$.
  \begin{equation}
  \frac{d\overline{I}}{d\sigma^2}
  =  \frac{\partial {\cal V}_0}{\partial \sigma^2} (\sigma^2,\bs{\dr},\bs{\dt})
  + \sum_{l=1}^L \frac{\partial {\cal V}_0}{\partial \kappa_l} (\sigma^2,\bs{\dr},\bs{\dt}) \cdot \frac{d\drl}{d\sigma^2}
  + \sum_{l=1}^L \frac{\partial {\cal V}_0}{\partial \tilde{\kappa}_l}(\sigma^2,\bs{\dr},\bs{\dt})\cdot \frac{d\dtl}{d\sigma^2}.
  \end{equation}
  It is straightforward to check that
  \begin{align}
    \left\{\begin{array}{l}
      \displaystyle
      \frac{\partial {\cal V}_0}{\partial \kappa_l}(\sigma^2,\bs{\kappa},\tilde{\bs{\kappa}})=-\sigma^2 t\big(\ftl(\bs{\kappa}, \I_t)-\tilde{\kappa}_l\big),
      \vspace{3mm}\\
      \displaystyle
      \frac{\partial {\cal V}_0}{\partial \tilde{\kappa}_l}(\sigma^2,\bs{\kappa},\tilde{\bs{\kappa}})=-\sigma^2t\big(\frl(\tilde{\bs{\kappa}})-\kappa_l\big).
    \end{array}\right.
      \label{eq:derivee-kappa}
  \end{align}
  Both partial derivatives are equal to zero at point $(\sigma^2,\bs{\dr}, \bs{\dt})$, as $(\bs{\dr}, \bs{\dt})$ verifies by definition (\ref{eq:canonique}) with $\Q=\I_t$.
  Therefore,
  \[
    \frac{d\overline{I}}{d\sigma^2}
    =  \frac{\partial {\cal V}_0}{\partial \sigma^2} (\sigma^2,\bs{\dr},\bs{\dt})=-t\sum_{l=1}^L\drl\dtl,
  \]
  which, together with Proposition \ref{prop:integrable}, leads to $I=\overline{I}+\mathcal{O}\left(\frac{1}{t}\right)$.
  
\end{proof}

\subsection{The approximation $\overline{I}({\bf Q})$}
We now consider the dependency in ${\bf Q}$ of the approximation $\bar{I}({\bf Q})$.
We previously considered the case ${\bf Q}={\bf I}$; to address the general case it is sufficient to change matrices $(\Ctl)_{l=1, \ldots, L}$ into  $({\bf Q}^{1/2}\Ctl {\bf Q}^{1/2})_{l=1, \ldots, L}$ in \ref{sec:canon-eq} and \ref{sec:appGauss}.
Hence the following Corollary of Theorem \ref{th:appr}:
\medskip
\begin{corollary}
 \label{th:apprQ}
 Assume that, for every $j \in \left\{ 1, \ldots, L\right\}$, $\sup_t\|\Crj\|<+\infty$, $\sup_t\|\Ctj\|<+\infty$,
 $\inf_t\left(\frac{1}{t}\mathrm{Tr\,}\Crj\right)>0$ and $\inf_t\lambda_{\min}(\Ctj)>0$.
 Then, for $\Q$ such as $\sup_t\| \Q \|<+\infty$,
 \[I(\Q)=\overline{I}(\Q)+\mathcal{O}\left(\frac{1}{t}\right).\]
\end{corollary}
Note that the technical assumptions on matrices $(\Ctl)_{l=1, \ldots, L}$ are slightly stronger than in Theorem \ref{th:appr} in order to ensure that
$\inf_t\left(\frac{1}{t}\mathrm{Tr\,}\left[\Q\Ctj\right]\right)>0$.
\medskip

We can now state an important result about the concavity of the function ${\bf Q} \mapsto \overline{I}({\bf Q})$, a result which will be highly needed for its optimization in section \ref{sec:algo}.
\medskip	
\begin{theorem}
 \label{th:concavs}
 ${\bf Q} \mapsto \overline{I}({\bf Q})$ is a strictly concave function over the compact set ${\cal C}_1$.
\end{theorem}
\medskip
\begin{proof}
  We here only prove the concavity of $\overline{I}({\bf Q})$.
  The proof of the strict concavity is quite tedious, but essentially the same as in \cite{dumont2010on} section IV (see also the extended version \cite{dumont-2007}). It is therefore omitted.
  
  Denote by $\otimes$ the Kronecker product of matrices. Let us introduce the following matrices:
  \[
     {\bs \Delta}^{(l)}={\bf I}_m \otimes \Crl, \tilde{\bs \Delta}^{(l)}={\bf I}_m \otimes \Ctl, \check{\bf Q}={\bf I}_m \otimes {\bf Q}.
  \]
  We now denote
  \[
    \check{\bf H}(z)=\sum_{l=1}^L\check{\H}^{(l)} z^{-(l-1)} \text{ with } \check{\H}^{(l)}=\frac{1}{\sqrt{mt}}({\bf \Delta^{(l)}})^{1/2}\check{\bf W}_l(\tilde{\bf \Delta}^{(l)})^{1/2},
  \]
  where $\check{\bf W}$ is a $rm \times tm$ matrix whose entries are independent and identically distributed complex circular Gaussian random variables with variance $1$.
  Introducing $I_m(\check{\bf Q})$ the ergodic mutual information associated with channel $\check{\bf H}(z)$:
  \[
    I_m(\check{\bf Q})=\mathbb{E}_{\check{\bf H}}\log\left|{\bf I}+\frac{{\bf \check{H} \check{Q} \check{H}}^H}{\sigma^2}\right|,
  \]
  where $\check{\bf H}=\check{\bf H}(1)=\sum_l \check{\H}^{(l)}$.
  Using the results of \cite{Moustakas-Simon-07} and Theorem \ref{th:appr}, it is clear that $I_m(\check{\bf Q})$ admits an asymptotic approximation $\bar{I}_m(\check{\bf Q})$.
  Due to the block-diagonal nature of matrices ${\bs \Delta}^{(l)}, \tilde{\bs \Delta}^{(l)}$ and $\check{\bf Q}$, it is straightforward to show that
  $\delta_l({\bf Q})=\delta_l(\check{\bf Q}), \tilde{\delta}_l({\bf Q})=\tilde{\delta}_l(\check{\bf Q})$
  and that, as a consequence,
  \[
    \frac{1}{m}\bar{I}_m(\check{\bf Q})= \bar{I}({\bf Q}),
  \]
  and thus
  \[
    \lim_{m\rightarrow\infty}\frac{1}{m}I_m(\check{\bf Q})=\bar{I}({\bf Q}).
  \]
  As $\check{\bf Q}\mapsto I_m(\check{\bf Q})$ is concave, we can conclude that $\bar{I}({\bf Q})$ is concave as a pointwise limit of concave functions.
\end{proof}
\medskip

As $\overline{I}(\Q)$ is strictly concave on $\mathcal{C}_1$ by Theorem \ref{th:concavs}, it admits a unique argmax that we denote $\overline{\Q}_*$.
We recall that $I(\Q)$ is strictly concave on $\mathcal{C}_1$ and that we denoted $\Q_*$ its argmax.
In order to clarify the relation between $\overline{\Q}_*$  and $\Q_*$ we introduce the following proposition which establishes that the maximization of $\overline{I}(\Q)$ is equivalent to the maximization of $I(\Q)$ over $\mathcal{C}_1$, up to a $\mathcal{O}\left( \frac{1}{t} \right)$ term.
\begin{proposition}
 Assume that, for every $j \in \left\{ 1, \ldots, L\right\}$,
 $\sup_t\|\Crj\|<+\infty$,
 $\sup_t\|\Ctj\|<+\infty$,
 $\inf_t\lambda_{\min}(\Crj)>0$
 and
 $\inf_t\lambda_{\min}(\Ctj)>0$.
 Then
 \[
  I(\overline{\Q}_*)=I(\Q_*)+\mathcal{O}\left(\frac{1}{t}\right).
 \]
 \label{prop:Q*-Qb*}
\end{proposition} \medskip
\begin{proof}
 The proof is very similar to the one of \cite[Proposition 3]{dumont2010on}.
 Assuming that
 $\sup_t\| \overline\Q_* \| < +\infty$
 and
 $\sup_t\| \Q_* \| < +\infty$
 we can apply Theorem \ref{th:apprQ} on $\overline\Q_*$ and $\Q_*$, hence
 \[
  \Big[I(\Q_*)- I(\overline \Q_*) \Big]
   + \Big[\overline I(\overline \Q_*)- \overline I( \Q_*) \Big]
  =   \Big[I(\Q_*)- \overline I( \Q_*) \Big]
   + \Big[\overline I(\overline \Q_*)- I(\overline \Q_*) \Big]
  = \mathcal{O}\left(\frac{1}{t}\right).
 \]
 Besides
 $I(\Q_*)- I(\overline \Q_*) \geq 0$
 and
 $\overline I(\overline \Q_*)- \overline I( \Q_*)\geq 0$,
 as
 $\Q_*$ and $\overline \Q_*$ respectively maximize $I(\Q)$ and $\overline I(\Q)$.
 Therefore $I(\Q_*)-I(\overline{\Q}_*)=\mathcal{O}\left(\frac{1}{t}\right)$.
 
 One can prove $\sup_t\| \overline\Q_* \|< +\infty$ using the same arguments as in \cite[Appendix III]{dumont2010on}.
 It essentially lies in the fact that $\overline\Q_*$ is the solution of a waterfilling algorithm, which will be shown independently from this result in next section (see Proposition \ref{prop:waterfilling}).

 Concerning $\sup_t\| \Q_* \|< +\infty$, the proof is identical to \cite[Appendix III]{dumont2010on}, 
 one just needs to replace
 $\sqrt{\frac{K}{K+1}}\A$ by $\frac{1}{\sqrt{t}}\sum_{l=2}^L\big(\Crl\big)^{1/2}\W_l\big(\Ctl\big)^{1/2}$
 and
 $\sqrt{\frac{1}{K+1}}\frac{1}{\sqrt{t}}\C_R^{1/2}\W\C_T^{1/2}$ by $\frac{1}{\sqrt{t}}\big(\Cr^{(1)}\big)^{1/2}\W_1\big(\Ct^{(1)}\big)^{1/2}$
 in the definition of $\H$.
 Then $S_j$, defined in \cite[(134)]{dumont2010on}, becomes
 \begin{equation}
  S_j=
  2\mathrm{Re}\left\{  \frac{1}{t} {\bf u}_{j}^{\perp H} \big(\Cr^{(1)}\big)^{1/2} \R_j \left(\sum_{l=2}^L \big(\Crl\big)^{1/2} \z_{l,j}+\big(\Cr^{(1)}\big)^{1/2}{\bf u}_{j} \right)  \right\}
  +
  \frac{1}{t}{\bf u}_{j}^{\perp H} \big(\Cr^{(1)}\big)^{1/2} \R_j \big(\Cr^{(1)}\big)^{1/2} {\bf u}_{j}^\perp,
    \label{eq:Sj}
 \end{equation}
 where
 $\R_j$ has the same definition as in \cite{dumont2010on},
 $\z_{l,j}$ is the $j^{th}$ column of matrix $\W_l\big(\Ctl\big)^{1/2}$
 and
 $\z_j=\z_{1,j}={\bf u}_{j}+{\bf u}_{j}^{\perp}$ with ${\bf u}_{j}$ the conditional expectation
 $
  {\bf u}_{j}
  =\mathbb{E}\left[ \z_{1,j} \Big| \left(\z_{1,k}\right)_{1\leq k \leq t, k\neq j} \right]
 $.
 As the vector ${\bf u}_j^\perp$ is independent from $\R_j$ and from $\z_{l,k}$, $k=1,\ldots,t$, $l=2,\ldots,L$
 we can easily prove that the first term of the right-hand side of \eqref{eq:Sj} is a $\mathcal{O}\left(\frac{1}{t}\right)$.
 The second term of the right-hand side of \eqref{eq:Sj} is moreover close from 
 $\rho_j=\frac{1}{t}\big[(\Ct^{(1)})^{-1}\big]_{jj}^{-1}\mathrm{Tr}\left(\R_j\Cr^{(1)}\right)$.
 In fact it is possible to prove that there exists a constant $C_1$ such that $\mathbb{E}\left[(S_j-\rho_j)^2\right]<\frac{C_1}{t}$
 (see \cite{dumont2010on} for more details).

 The rest of the proof of \cite[Proposition 3 (ii)]{dumont2010on} can then follow.

\end{proof}

\section{Maximization algorithm}
\label{sec:algo}
 Proposition \ref{prop:Q*-Qb*} shows that it is relevant to maximize $\overline{I}(\Q)$ over $\mathcal{C}_1$.
 In this section we propose a maximization algorithm for the large system approximation $\overline I(\Q)$.
 We first introduce some classical concepts and results needed for the optimization of ${\bf Q}\mapsto\overline{I}({\bf Q})$. \medskip
\begin{definition}
 Let $\phi$ be a function defined on the convex set $\mathcal{C}_1$. Let ${\bf P},{\bf Q} \in \mathcal{C}_1$.
 Then $\phi$ is said to be differentiable in the G\^ateaux sense (or G\^ateaux differentiable) at point ${\bf Q}$ in the direction ${\bf P}-{\bf Q}$ if the following limit exists:
 \[
  \lim_{\lambda\rightarrow 0^+} \frac{\phi({\bf Q}+\lambda({\bf P}-{\bf Q}))-\phi({\bf Q})}{\lambda}.
 \]
In this case, this limit is noted $\langle \phi'({\bf Q}),{\bf P}-{\bf Q}\rangle$.
\end{definition} \medskip
Note that $\phi({\bf Q}+\lambda({\bf P}-{\bf Q}))$ makes sense for $\lambda\in[0,1]$, as ${\bf Q}+\lambda({\bf P}-{\bf Q})=(1-\lambda){\bf Q}+\lambda{\bf P}$ naturally belongs to $\mathcal{C}_1$. We now establish the following result:
\medskip
\begin{proposition}
 Then, for each $\P,\Q\in \mathcal{C}_1$,
 functions $\Q \mapsto \drl(\Q)$, $\Q \mapsto \dtl(\Q)$, $l=1,\ldots,L$, as well as function $\Q \mapsto \overline{I}(\Q)$ are G\^ateaux differentiable at $\Q$ in the direction $\P-\Q$.
 \label{prop:diff_Ib}
\end{proposition}\medskip
\begin{proof}
 See Appendix \ref{proof:diff}.
\end{proof}\medskip

In order to characterize matrix $\overline{\Q}_*$ we recall the following result:

\begin{proposition}
 Let $\phi:\mathcal{C}_1\rightarrow\mathbb{R}$ be a strictly concave function.
 Then,
 \begin{enumerate}[(i)]
  \item $\phi$ is G\^ateaux differentiable at ${\bf Q}$ in the direction ${\bf P}-{\bf Q}$ for each ${\bf P}, {\bf Q} \in \mathcal{C}_1$,
  \item ${\bf Q}_{opt}$ is the unique argmax of $\phi$ on $\mathcal{C}_1$ if and only if it verifies:
  \begin{equation}
   \label{eq:carac-max}
   \forall {\bf Q} \in \mathcal{C}_1, \, \langle \phi'({\bf Q}_{opt}),{\bf Q}-{\bf Q}_{opt}\rangle\leq0.
  \end{equation}
 \end{enumerate}
\end{proposition}
This proposition is standard (see for example Chapter 2 of \cite{borwein2006convex}).

In order to introduce our maximization algorithm, we consider the function ${\cal V}({\bf Q},{\bs \kappa},\tilde{\bs \kappa})$ defined by:
\begin{equation}
   {\cal V}({\bf Q},{\bs \kappa},\tilde{\bs \kappa}) = \log | {\bf I} + {\bf C}(\tilde{\bs \kappa}) |
   +\log |{\bf I} + {\bf Q} \tilde{\bf C}({\bs \kappa}) |
   -\sigma^{2} t \sum_{l=1}^{L}  \kappa_l \tilde\kappa_l.
   \label{def:V}
\end{equation}
We recall that
$\tilde{\bf C}({\bs \kappa})=\sum_{l=1}^{L} \kappa_l \Ctl$ and ${\bf C}(\tilde{\bs \kappa})=\sum_{l=1}^{L} \tilde\kappa_l \Crl$. 
Note that we have ${\cal V}({\bf Q},{\bs \delta}({\bf Q}),\tilde{\bs \delta}({\bf Q}))=\overline{I}({\bf Q})$. We have then the following result:\medskip
\begin{proposition}
\label{prop:waterfilling}
 Denote by $\bs{\delta}_*$ and $\tilde{\bs{\delta}}_*$ the quantities ${\bs \delta}(\overline{\bf Q}_*)$ and $\tilde{\bs \delta}(\overline{\bf Q}_*)$.
 Matrix $\overline{\bf Q}_*$ is the solution of the standard waterfilling problem:
 maximize over ${\bf Q \in {\cal C}_1}$ the function $\log|{\bf I}+{\bf Q}\tilde{\bf C}(\bs{\delta}_*)|$.
\end{proposition} \medskip
\begin{proof}
We first remark that maximizing function ${\bf Q}\mapsto\log|{\bf I}+{\bf Q}\tilde{\bf C}(\bs{\delta}_*)|$ is equivalent to maximizing function ${\bf Q}\mapsto{\cal V}({\bf Q},\bs{\delta}_*,\tilde{\bs{\delta}}_*)$ by (\ref{def:V}).
The proof then relies on the observation hereafter proven that, for each ${\bf P} \in {\cal C}_1$,
\begin{equation}
 \langle\overline{I}'(\overline{{\bf Q}}_*), {\bf P} - \overline{{\bf Q}}_*\rangle
 = \langle \mathcal{V}' (\overline{{\bf Q}}_*,\bs{\delta}_*,\tilde{\bs{\delta}}_*), {\bf P} - \overline{{\bf Q}}_*\rangle,
 \label{eq:nablas}
\end{equation}
where $\langle \mathcal{V}' (\overline{{\bf Q}}_*,\bs{\delta}_*,\tilde{\bs{\delta}}_*), {\bf P} - \overline{{\bf Q}}_*\rangle$ is the G\^ateaux differential of function ${\bf Q}\mapsto\mathcal{V} (\Q,\bs{\delta}_*,\tilde{\bs{\delta}_*})$ at point $\overline{{\bf Q}}_*$ in direction ${\bf P} - \overline{{\bf Q}}_*$.
Assuming (\ref{eq:nablas}) is verified, (\ref{eq:carac-max}) yields that 
$\langle\mathcal{V'} (\overline{{\bf Q}}_*,\bs{\delta}_*,\tilde{\bs{\delta}}_*), {\bf P} - \overline{{\bf Q}}_*\rangle \leq 0$
for each matrix ${\bf P} \in \mathcal{C}_1$.
And as the function ${\bf Q}\mapsto \mathcal{V}({\bf Q},\bs{\delta}_*,\tilde{\bs{\delta}}_*)$ is strictly concave on $\mathcal{C}_1$, its unique argmax on ${\cal C}_1$ coincides with
$\overline{{\bf Q}}_*$.

It now remains to prove (\ref{eq:nablas}). Consider ${\bf P}$ and ${\bf Q}$ $\in \mathcal{C}_1$. Then,
\begin{align}
 \langle\overline{I}'({\bf Q}), {\bf P} - {\bf Q}\rangle
 = &\langle \mathcal{V}' ({\bf Q},\bs{\delta}({\bf Q}),\tilde{\bs{\delta}}({\bf Q})), {\bf P} - {\bf Q}\rangle
 + \sum_{l=1}^L\frac{\partial {\cal V}}{\partial \kappa_l}({\bf Q},\bs{\delta}({\bf Q}),\tilde{\bs{\delta}}({\bf Q}))\langle\delta_l'({\bf Q}) , {\bf P} - {\bf Q}\rangle \notag
 \\& + \sum_{l=1}^L\frac{\partial {\cal V}}{\partial \tilde{\kappa}_l}({\bf Q},\bs{\delta}({\bf Q}),\tilde{\bs{\delta}}({\bf Q}))\langle \tilde{\delta}_l'({\bf Q}) , {\bf P} - {\bf Q}\rangle. \label{eq:derivpar}
\end{align}
Similarly to \eqref{eq:derivee-kappa}, partial derivatives
$\frac{\partial {\cal V}}{\partial \kappa_l}({\bf Q},\bs{\kappa},\tilde{\bs{\kappa}})=-\sigma^2t\big(\ftl(\bs{\kappa}, {\bf Q})-\tilde{\kappa}_l\big)$
and
$\frac{\partial {\cal V}}{\partial \kappa_l}({\bf Q},\bs{\kappa},\tilde{\bs{\kappa}})=-\sigma^2t\big(\frl(\tilde{\bs{\kappa}})-\kappa_l\big)$
are equal to zero at point
$({\bf Q},\bs{\delta}({\bf Q}), \tilde{\bs{\delta}}({\bf Q}))$,
as $(\bs{\delta}({\bf Q}), \tilde{\bs{\delta}}({\bf Q}))$ verifies (\ref{eq:canonique}) by definition.
Therefore, letting ${\bf Q}=\overline{\bf Q}_*$ in (\ref{eq:derivpar}) yields:
\[
  \langle \overline{I}'(\overline{{\bf Q}}_*), {\bf P} - \overline{{\bf Q}}_*\rangle
 = \langle \mathcal{V}' (\overline{{\bf Q}}_*,\bs{\delta}(\overline{{\bf Q}}_*),\tilde{\bs{\delta}}(\overline{{\bf Q}}_*)), {\bf P} - \overline{{\bf Q}}_*\rangle.
\]
\end{proof}

Proposition \ref{prop:waterfilling} shows that the optimum matrix is solution of a waterfilling problem associated to 
the covariance matrix $\tilde{\bf C}(\bs{\delta}_*)$. This result cannot be used to evaluate $\overline{{\bf Q}}_{*}$, because the matrix  $\tilde{\bf C}(\bs{\delta}_*)$ itself depends of $\overline{{\bf Q}}_{*}$. 
However, it provides some insight on the structure of the optimum matrix: the eigenvectors of $\overline{{\bf Q}}_{*}$ coincide with the eigenvectors of a linear combination of matrices $\Ctl$, the $\drl({\bf Q}_*)$ being the coefficients of this linear combination. This is in line with the result of \cite{Moustakas-Simon-07} Appendix VI.

We now introduce our iterative algorithm for optimizing $\overline{I}({\bf Q})$:
\begin{itemize}
	\item Initialization: ${\bf Q}_0={\bf I}$.
	\item Evaluation of ${\bf Q}_k$ from ${\bf Q}_{k-1}$:
	$({\bs{\delta}}^{(k)}, \tilde{\bs{\delta}}^{(k)})$ is defined as the unique solution of (\ref{eq:canonique}) in which ${\bf Q}={\bf Q}_{k-1}$.
	Then ${\bf Q}_k$ is defined as the maximum of function ${\bf Q}\mapsto\log\left|{\bf I}+{\bf Q}\tilde{\bf C}(\bs{\delta}^{(k)})\right|$ on ${\cal C}_1$.
\end{itemize}

\medskip
We now establish a result which implies that, if the algorithm converges, then it converges towards the optimal covariance matrix $\overline{{\bf Q}}_*$. 

\begin{proposition}
\label{prop:convergence}
Assume that 
\begin{equation}
  \label{eq:hypothese}
  \lim_{k \rightarrow \infty} {\bs{\delta}}^{(k)} - {\bs{\delta}}^{(k-1)} =\lim_{k \rightarrow \infty} {\tilde{\bs{\delta}}}^{(k)} - \tilde{{\bs{\delta}}}^{(k-1)} =  0.
\end{equation} 
Then, the algorithm converges towards matrix $\overline{{\bf Q}}_*$. 
\end{proposition}
\medskip
\begin{proof}
The sequence $({\bf Q}_k)$ belongs to the set ${\cal C}_1$.
As ${\cal C}_1$ is compact, we just have to verify that every convergent subsequence $({\bf Q}_{\psi(k)})_{k\in\mathbb{N}}$ extracted from $({\bf Q}_k)_{k\in\mathbb{N}}$ converges towards 
$\overline{{\bf Q}}_*$.
For this, we denote by $\overline{{\bf Q}}_{\psi,*}$ the limit of the above subsequence, and prove that this matrix verifies property (\ref{eq:carac-max}) with $\phi=\overline{I}$.
Vectors $\bs{\delta}^{\psi(k)+1}$ and $\tilde{\bs{\delta}}^{\psi(k)+1}$ are defined as the solutions of (\ref{eq:canonique}) with ${\bf Q}={\bf Q}_{\psi(k)}$.
Hence, due to the continuity of functions ${\bf Q}\mapsto\drl({\bf Q})$ and ${\bf Q} \mapsto \dtl({\bf Q})$, sequences $\Big(\bs{\delta}^{\psi(k)+1}\Big)_{k\in\mathbb{N}}$ and $\Big(\tilde{\bs{\delta}}^{\psi(k)+1}\Big)_{k\in\mathbb{N}}$ converge towards $\bs{\delta}^{\psi,*}=\bs{\delta}(\overline{\bf Q}_{\psi,*})$ and $\tilde{\bs{\delta}}^{\psi,*}=\tilde{\bs{\delta}}(\overline{\bf Q}_{\psi,*})$ respectively.
Moreover, $\Big(\bs{\delta}^{\psi,*}, \tilde{\bs{\delta}}^{\psi,*}\Big)$ is solution of system (\ref{eq:canonique}) in which matrix ${\bf Q}$ coincides with $\overline{{\bf Q}}_{\psi,*}$.
Therefore,
\[
 \frac{\partial {\cal V}}{\partial \kappa_l}\left(\overline{{\bf Q}}_{\psi,*},\bs{\delta}^{\psi,*},\tilde{\bs{\delta}}^{\psi,*}\right)
 =\frac{\partial {\cal V}}{\partial \tilde{\kappa}_l}\left(\overline{{\bf Q}}_{\psi,*},\bs{\delta}^{\psi,*},\tilde{\bs{\delta}}^{\psi,*}\right)
 = 0.
\]
As in the proof of Proposition \ref{prop:waterfilling}, this leads to
\begin{equation}
 \langle \overline{I}'(\overline{{\bf Q}}_{\psi,*}), {\bf P} - \overline{{\bf Q}}_{\psi,*}\rangle
 =\langle {\cal V}'(\overline{{\bf Q}}_{\psi,*}, \bs{\delta}_{\psi,*}, \tilde{{\bs \delta}}_{\psi,*}), {\bf P} - \overline{{\bf Q}}_{\psi,*}\rangle
  \label{eq:conv_proof}
\end{equation}
for every ${\bf P} \in \mathcal{C}_1$.
It remains to show that the right-hand side of \eqref{eq:conv_proof} is negative to complete the proof. For this, we use that 
${\bf Q}_{\psi(k)}$ is the argmax over ${\cal C}_1$ of function ${\bf Q} \mapsto {\cal V}\Big({\bf Q}, \bs{\delta}^{\psi(k)}, \tilde{\bs{\delta}}^{\psi(k)}\Big)$.
Therefore, 
\begin{equation}
 \label{eq:derivee-negative}
 \langle {\cal V}'({\bf Q}_{\psi(k)}, \bs{\delta}_{\psi(k)}, \tilde{{\bs \delta}}_{\psi(k)}), {\bf P} - {\bf Q}_{\psi(k)}\rangle 
 \leq 0
 \ \ \forall \ \P \in\mathcal{C}_1.
\end{equation}
By condition (\ref{eq:hypothese}), sequences $(\bs{\delta}_{\psi(k)})$ and $(\tilde{\bs{\delta}}_{\psi(k)})$ also converge towards 
$\bs{\delta}^{\psi,*}$ and  $\tilde{\bs{\delta}}^{\psi,*}$ respectively. Taking the limit of (\ref{eq:derivee-negative}) 
when $k\rightarrow\infty$ eventually shows that $\langle {\cal V}'(\overline{{\bf Q}}_{\psi,*}, \bs{\delta}_{\psi,*}, \tilde{{\bs \delta}}_{\psi,*}), {\bf P} - \overline{{\bf Q}}_{\psi,*}\rangle \leq 0$ as required.
\end{proof}\medskip

To conclude, if the algorithm is convergent, that is, if the sequence of $({\bf Q}_k)_{k\in\mathbb{N}}$ converges towards a certain matrix, then the $\drl^{(k)}=\drl({\bf Q}_{k-1})$ and the $\dtl^{(k)}=\dtl({\bf Q}_{k-1})$ converge as well when $k\rightarrow\infty$.
Condition (\ref{eq:hypothese}) is then verified, hence, 
if the algorithm is convergent, it converges towards $\overline{\bf Q}_*$.
Although the convergence of the algorithm has not been proved, this result is encouraging and suggests that the algorithm is reliable.
In particular, in all the conducted simulations the algorithm was converging.
In any case, condition (\ref{eq:hypothese}) can be easily checked.
If it is not satisfied, it is possible to modify the initial point ${\bf Q}_0$ as many times as needed to ensure the convergence.

\section{Numerical Results}
\label{sec:simulations}

We provide here some simulations results to evaluate the performance of the proposed approach. We use the propagation model introduced in \cite{boelcskei}, in which each path corresponds to a scatterer cluster characterized by a mean angle of departure, a mean angle of arrival and an angle spread for each of these two angles.

\begin{figure}[!t]
	\centering
	\subfigure[$r=t=4$ \label{fig_comp}]{\includegraphics[width=.49\textwidth]{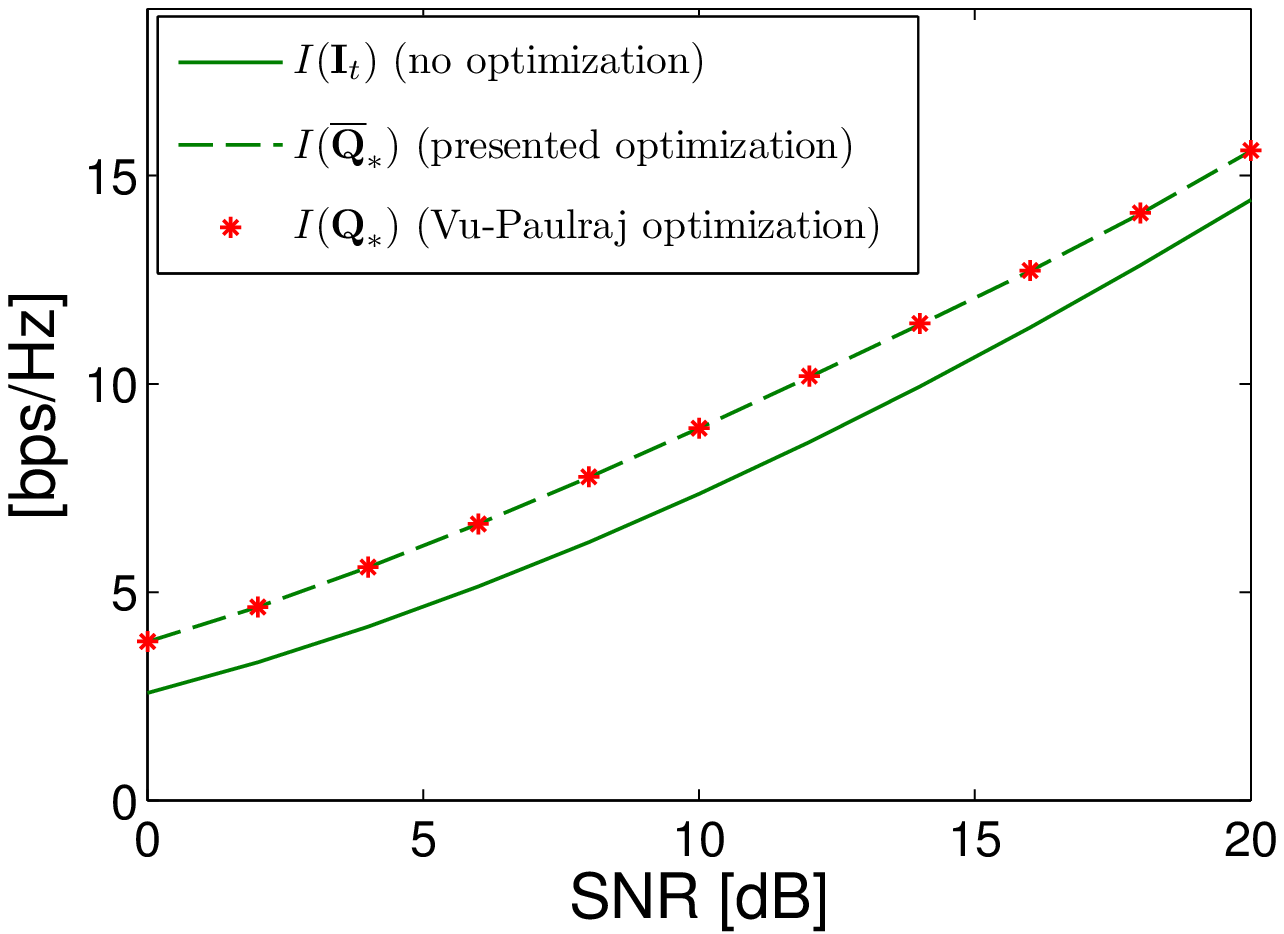}}
	\subfigure[$r=t=8$ \label{fig_comp2}]{\includegraphics[width=.49\textwidth]{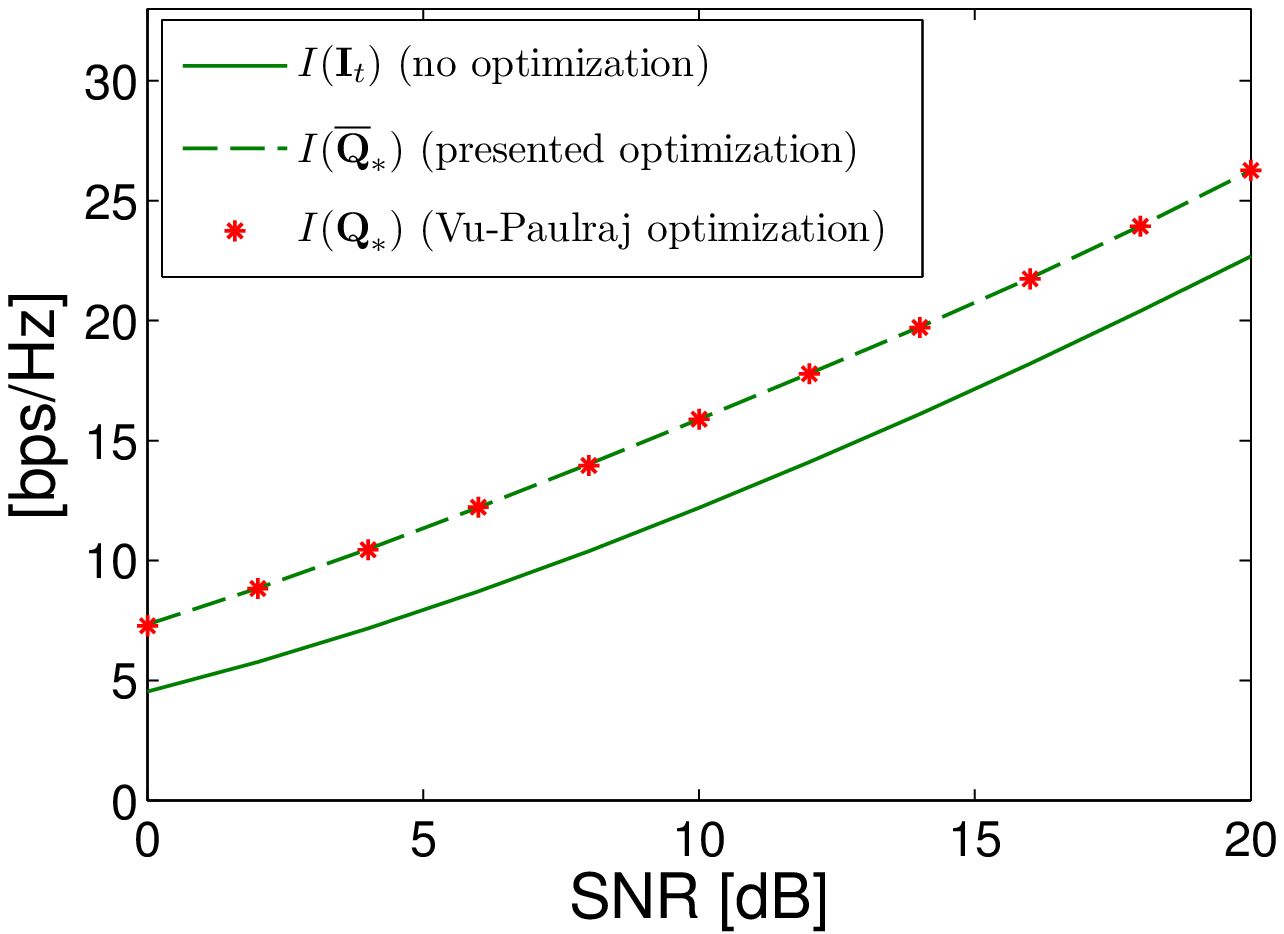}}
	\caption{Comparison with Vu-Paulraj algorithm}
	\label{fig_subf}
	\end{figure}

In the featured simulations for Fig. \ref{fig_comp} (respectively Fig. \ref{fig_comp2}), we consider a frequency selective MIMO system with $r=t=4$ (respectively $r=t=8$), a carrier frequency of 2GHz, a number of paths $L=5$.
The paths share the same power, and their mean departure angles and angles spreads are given in Table~\ref{table_para} in radians.
In both Fig. \ref{fig_comp} and \ref{fig_comp2}, we have represented the EMI $I(\I_t)$ (i.e. without optimization), and the optimized EMI $I(\overline{\bf Q}_*)$ (i.e. with an input covariance matrix maximizing the approximation $\overline{I}$).
The EMI are evaluated by Monte-Carlo simulations, with $10^5$ channel realizations.
The EMI optimized with Vu-Paulraj algorithm \cite{Vu-Paulraj-05} is also represented for comparison.

Vu-Paulraj's algorithm is composed of two nested iterative loops.
The inner loop evaluates
$\Q^{(n)}_*={\mathrm{argmax}}\left\{ I(\Q) + k_{\mathrm{barrier}} \log|\Q| \right\}$
thanks to the Newton algorithm with the constraint $\frac{1}{t}\mathrm{Tr}\,\Q= 1$, for a given value of $k_{\mathrm{barrier}}$ and a given starting point $\Q^{(n)}_0$.
Maximizing $I(\Q) + k_{\mathrm{barrier}} \log|\Q|$ instead of $I(\Q)$ ensures that $\Q$ remains positive semi-definite through the steps of the Newton algorithm; this is the so-called barrier interior-point method.
The outer loop then decreases $k_{\mathrm{barrier}}$ by a certain constant factor $\mu$ and gives the inner loop the next starting point $\Q^{(n+1)}_0=\Q^{(n)}_*$.
The algorithm stops when the desired precision is obtained, or, as the Newton algorithm requires heavy Monte-Carlo simulations for the evaluation of the gradient and of the Hessian of $I({\bf Q})$, when the number of iterations of the outer loop reaches a given number $N_{\max}$.
As in \cite{Vu-Paulraj-05} we took $N_{\max}=10$, $\mu=100$, $2\cdot10^4$ trials for the Monte-Carlo simulations, and we started with $k_{\mathrm{barrier}}=\frac{1}{100}$.

Both Fig. \ref{fig_comp} and \ref{fig_comp2} show that maximizing $\overline{I}({\bf Q})$ over the input covariance leads to significant improvement for $I({\bf Q})$.
Our approach provides the same results as Vu-Paulraj's algorithm.
Moreover our algorithm is computationally much more efficient: in Vu-Paulraj's algorithm, the evaluation of the gradient and of the Hessian of $I({\bf Q})$ needs heavy Monte-Carlo simulations. Table \ref{table_comp} gives for both algorithms the average execution time in seconds to obtain the input covariance matrix, on a 3.16GHz Intel Xeon CPU with 8GB of RAM, for a number of paths $L=3$, $L=4$ and $L=5$, given $r=t=4$.

\begin{table}[!t]
\renewcommand{\arraystretch}{1.3}
\caption{Paths angular parameters \normalfont{\scriptsize\it(in radians)}}
\label{table_para}
\centering
\begin{tabular}{|c|c|c|c|c|c|}
\cline{2-6}\multicolumn{1}{c|}{}  & $l=1$ & $l=2$ & $l=3$ & $l=4$ & $l=5$
\\\hline {mean departure angle} &$6.15$ & $3.52$ & $4.04$ & $2.58$ & $2.66$
\\\hline {departure angle spread} & $0.06$ & $0.09$ & $0.05$ & $0.05$ & $0.03$
\\\hline {mean arrival angle} &$4.85$ & $3.48$ & $1.71$ & $5.31$ & $0.06$
\\\hline {arrival angle spread} &$0.06$ &$0.08$ & $0.05$ & $0.02$ & $0.11$
\\\hline
\end{tabular}
\end{table}

\begin{table}[!t]
\renewcommand{\arraystretch}{1.3}
\caption{Average execution time \normalfont{\scriptsize\it(in seconds)}}
\label{table_comp}
\centering
\begin{tabular}{|c|c|c|c|}
\cline{2-4}\multicolumn{1}{c|}{} & $L=3$& $L=4$& $L=5$\\
\hline
Vu-Paulraj & $681$& $884$& $1077$\\
\hline
New algorithm & $7.0\cdot10^{-3}$& $7.4\cdot10^{-3}$& $8.3\cdot10^{-3}$\\
\hline
\end{tabular}
\end{table}

\section{Conclusion}
In this paper we have addressed the evaluation of the capacity achieving covariance matrices of frequency selective MIMO channels.
We have first clarified the definition of the large system approximation of the EMI and rigorously proved its expression and convergence speed with Gaussian methods.
We have then proposed to optimize the EMI through this approximation, and have introduced an attractive iterative algorithm based on an iterative waterfilling scheme.
Numerical results have shown that our approach provides the same results as a direct approach, but in a more efficient way in terms of computation time.




\appendices

\section{Proof of the existence of a solution}
\label{apx:delta_exist-unique}

To study (\ref{eq:canonique}), it is quite useful to interpret functions $\frl$ and $\ftl$ as functions of the parameter $-\sigma^{2} \in \mathbb{R}^{-}$,  to extend 
their domain of validity from $\mathbb{R}^{-}$ to  $\mathbb{C}-\mathbb{R}^{+}$, and to use powerful results concerning certain class of analytic functions. We therefore define the functions $g(\bs{\tilde\psi})(z)$ and $g(\bs\psi)(z)$ as
\begin{align}
	&g(\bs{\tilde{\psi}})(z)=\left[\begin{array}{c}g_1(\bs{\tilde{\psi}})(z)\\...\\g_L(\bs{\tilde{\psi}})(z)\end{array}\right] \text{  where } g_l(\bs{\tilde{\psi}})(z)=\frac{1}{t} \mathrm{Tr} \left[ \Crl {\bf T}^{\bs{\tilde{\psi}}}(z) \right], \notag
	\\
	&\tilde{g}(\bs{\psi})(z)=\left[\begin{array}{c}\tilde{g_1}(\bs{\psi})(z)\\...\\\tilde{g_L}(\bs{\psi})(z)\end{array}\right] \text{  where } \tilde{g_l}(\bs{\psi})(z)=\frac{1}{t} \mathrm{Tr} \left[ \Ctl \tilde{\bf{T}}^{\bs{\psi}}(z) \right], \notag
\end{align}
with 
$\bs{\psi}(z)=(\psi_1(z), ... , \psi_L(z))^T$,
$\bs{\tilde{\psi}}(z)=(\tilde{\psi_1}(z) , ... , \tilde{\psi_L}(z))^T$
and where matrices ${\bf T}^{\bs{\tilde{\psi}}}(z)$ and $\tilde{{\bf T}}^{\bs{\psi}}(z)$ are defined by
\begin{align}
  &{\bf T}^{\bs{\tilde{\psi}}}(z) =\bigg[-z \bigg( {\bf I} + \sum_{j=1}^{L} \tilde{\psi}_j(z) \Crj \bigg)\bigg]^{-1},
  \label{eq:Tz}
  \\
  &\tilde{{\bf T}}^{\bs{\psi}}(z)= \bigg[-z \bigg( {\bf I} + \sum_{j=1}^{L} \psi_j(z) \Ctj \bigg)\bigg]^{-1}.
  \label{eq:Ttz}
\end{align}
In order to explain the following results, we now have to introduce the concept of Stieltj\`es transforms.
\medskip

\begin{definition}
\label{def:stieljes}
  Let $\mu$ be a finite\footnote{finite means that $\mu(\mathbb{R}^{+}) < \infty$} positive measure carried by $\mathbb{R}^{+}$. The Stieltj\`es transform of
  $\mu$ is the function $s(z)$ defined for  $z \in \mathbb{C}-\mathbb{R}^{+}$ by
  \begin{equation}
    \label{eq:def-stieljes}
    s(z) = \int_{\mathbb{R}^{+}} \frac{d \mu(\lambda)}{\lambda - z}.
  \end{equation}
\end{definition} \medskip
In the following, the class of all Stieltj\`es transforms of finite positive measures carried by $\mathbb{R}^{+}$ is denoted ${\cal S}(\mathbb{R}^{+})$.
We now state some of the properties of the elements of $\mathcal{S}(\mathbb{R}^+)$. \medskip

\begin{proposition}
\label{pr:stieljes}
Let $s(z)\in {\cal S}(\mathbb{R}^{+})$, and $\mu$ its associated measure. Then we have the following results:
\begin{enumerate}[(i)]
 \item $s(z)$ is analytic on  $\mathbb{C}-\mathbb{R}^{+}$,
 \item $\mathrm{Im}(s(z)) > 0$ if $\mathrm{Im}(z) > 0$, and $\mathrm{Im}(s(z)) < 0$ if $\mathrm{Im}(z) < 0$,
 \item $\mathrm{Im}(zs(z)) > 0$ if $\mathrm{Im}(z) > 0$, and $\mathrm{Im}(zs(z)) < 0$ if $\mathrm{Im}(z) < 0$,
 \item $s(-\sigma^{2}) > 0$ for $\sigma^{2}>0$,
 \item \label{pr:st-bounded} $|s(z)|\leq\frac{\mu(\mathbb{R}^+)}{d(z,\mathbb{R}^+)}$ for $z \in \mathbb{C-R}^+$,
 \item $\displaystyle \mu(\mathbb{R}^{+}) = \lim_{y \rightarrow \infty} -iy \, s(iy)$.
\end{enumerate}
\end{proposition} \medskip
\begin{proof}
 All the stated properties are standard material, see e.g. Appendix of \cite{krein1977markov}.
\end{proof} \medskip

Conversely, a useful tool to prove that a certain function belongs to ${\cal S}(\mathbb{R}^{+})$ is the following proposition: 
\begin{proposition}
  \label{pro:stieljes}
  Let $s$ be a function holomorphic on $\mathbb{C-R}^{+}$ which verifies the three following properties
  \begin{enumerate}[(i)]
  \item $\mathrm{Im}(s(z))>0$ if $\mathrm{Im}(z)>0$,
  \item $\mathrm{Im}(z s(z))>0$ if $\mathrm{Im}(z)>0$,
  \item $\displaystyle \sup_{y > 0}\left|iy \, s(iy)\right|<\infty$.
  \end{enumerate}
  Then $s\in{\cal S}(\mathbb{R}^+)$, and if $\mu$ represents the corresponding positive measure, then $\mu(\mathbb{R}^{+}) = \displaystyle\lim_{y \rightarrow \infty}(-iy \, s(iy))$.
\end{proposition} \medskip
\begin{proof}
 see Appendix of \cite{krein1977markov}.
\end{proof} \medskip

Now that we have recalled the notion of Stieltj\`es transforms and its associated basic properties we can introduce the following proposition: \medskip
\begin{proposition}
 \label{pr:iteration}
 Let $(\psi_l,\tilde \psi_l)_{l=1,\ldots,L} \in \mathcal{S}(\mathbb{R^+})$.
 We define functions $\varphi_l(z)$ and $\tilde \varphi_l(z)$, $l=1,\ldots,L$, as
 \[
  \left\{\begin{array}{l}
  \varphi_l(z)= \frac{1}{t}\mathrm{Tr}\left[\Crl{\bf T}^{\tilde{\bs{\psi}}}(z)\right],
  \vspace{1mm}\\
  \tilde \varphi_l(z)= \frac{1}{t}\mathrm{Tr}\left[\Ctl \tilde{\bf T}^{\bs{\psi}}(z)\right].
  \end{array}\right.
 \]
 Then we have the following results
 \begin{enumerate}[(i)]
  \item \label{pr:it-holo} ${\bf T}^{\tilde{\bs{\psi}}}$, $\tilde{\bf T}^{\bs{\psi}}$ are holomorphic on $\mathbb{C-R^+}$,
  \item \label{pr:it-ineq} $\|{\bf T}^{\tilde{\bs{\psi}}}(z)\|\leq\frac{1}{d(z,\mathbb{R}^+)}$ and $\|\tilde{\bf T}^{\bs{\psi}}(z)\|\leq\frac{1}{d(z,\mathbb{R}^+)}$ on $\mathbb{C-R^+}$,
  \item \label{pr:it-stable} $\varphi_l \in \mathcal{S}(\mathbb{R}^+)$ with the corresponding mass $\mu_l$ verifying $\mu_l(\mathbb{R}^+)=\frac{1}{t}\mathrm{Tr}\,\Crl$, and $\tilde \varphi_l \in \mathcal{S}(\mathbb{R}^+)$ with the corresponding mass $\tilde \mu_l$ verifying $\tilde \mu_l(\mathbb{R}^+)=\frac{1}{t}\mathrm{Tr}\,\Ctl$.
 \end{enumerate}
\end{proposition}\medskip

\begin{proof}
 For item (\ref{pr:it-holo}) we only have to check that
 $z \left({\bf I}+\sum_{j=1}^L\tilde \psi_j(z)\Crj \right)$
 is invertible for every $z \in \mathbb{C-R^+}$ to prove that ${\bf T}^{\tilde{\bs{\psi}}}$ is holomorphic on $\mathbb{C-R^+}$.
 The key point is to notice that, for any vector ${\bf v}$, for $z$ such that $\mathrm{Im}(z)>0$,
 \[
  \mathrm{Im}\Big\{{\bf v}^H z \Big({\bf I}+\sum_{j=1}^L\tilde \psi_j(z)\Crj \Big){\bf v}\Big\}=
  \mathrm{Im} \left\{z \right\} {\bf v}^H{\bf v}+\sum_{j=1}^L\mathrm{Im}\left\{z\tilde \psi_j(z) \right\}{\bf v}^H\Crj{\bf v}
  > 0.
 \]
 A similar inequality holds for $\mathrm{Im}(z)<0$, and the case $z \in \mathbb{R}^-$ is straightforward.
 
 Item (\ref{pr:it-stable}) can easily be proved thanks to Proposition \ref{pro:stieljes}.
 
 As for item (\ref{pr:it-ineq}), the proof is essentially the same as the proof of Proposition 5.1 item 3 in \cite{HLN07}, and is therefore omitted.
\end{proof}\medskip

We consider the following iterative scheme:
\begin{equation}
 \left\{\begin{array}{l}
    \bs{\psi}^{(n+1)}(z)=g(\bs{\tilde\psi}^{(n)})(z),
    \\
    \bs{\tilde\psi}^{(n+1)}(z)=\tilde g(\bs{\psi}^{(n)})(z),
 \end{array}\right.
 \label{ite-seq-stj}
\end{equation}
with a starting point $(\bs{\psi}^{(0)}(z),{\bs{\tilde\psi}^{(0)}}(z))$ in $\left({\cal S}(\mathbb{R}^{+})\right)^{2L}$.
Item (\ref{pr:it-stable}) of Proposition \ref{pr:iteration} then ensures that, for each $n\geq1$, $\bs{\psi}^{(n)}(z)$ and $\bs{\tilde\psi}^{(n)}(z)$ belong to $(\mathcal{S}(\mathbb{R}^+))^L$.
Moreover, 
\begin{align}
  \left|(\psi_l^{(n+1)} - \psi_l^{(n)})(z)\right|
  &= \left|g_l(\bs{\psi}^{(n)})(z)-g_l(\bs{\psi}^{(n-1)})(z)\right| \notag
  \\
  &= \frac{1}{t}\left|\mathrm{Tr}\left[\Crl({\bf T}^{(n)}(z)-{\bf T}^{(n-1)}(z))\right]\right|,
  \label{f_diff}
\end{align}
where matrices $\T^{(n)}(z)$ and $\tilde\T^{(n)}(z)$ are defined by $\T^{(n)}(z)=\T^{\bs{\tilde\psi}^{(n)}}(z)$, $\tilde\T^{(n)}(z)=\tilde\T^{\bs{\psi}^{(n)}}(z)$.
Using the equality ${\bf A}-{\bf B}={\bf A}\left({\bf B}^{-1}-{\bf A}^{-1}\right){\bf B}$, we then obtain
\begin{equation}
 {\bf T}^{(n)}(z)-{\bf T}^{(n-1)}(z) = 
 {\bf T}^{(n)}(z)\bigg(-z\sum_{j=1}^L\left(\tilde\psi^{(n-1)}_j(z)-\tilde\psi^{(n)}_j(z)\right)\Crj\bigg){\bf T}^{(n-1)}(z).
 \label{T_diff}
\end{equation}
Using (\ref{T_diff}) in (\ref{f_diff}) then yields
\begin{align}
  \left|(\psi_l^{(n+1)} - \psi_l^{(n)})(z)\right|
  &= \frac{|z|}{t}\bigg|\sum_{j=1}^L\left(\tilde\psi^{(n-1)}_j-\tilde\psi^{(n)}_j\right)(z)
  \mathrm{Tr} \left[\Crl{\bf T}^{(n)}(z)\Crj{\bf T}^{(n-1)}(z)\right] \bigg| \notag
  \\
  &\leq \frac{|z|}{t}\sum_{j=1}^L\left|\left(\tilde\psi^{(n-1)}_j-\tilde\psi^{(n)}_j\right)(z)\right|
  \left|\mathrm{Tr} \left[\Crl{\bf T}^{(n)}(z)\Crj{\bf T}^{(n-1)}(z)\right]\right|.
\end{align}
The trace in the above expression can be bounded with the help of $\Cmax=\max_{l,j}\left\{\| \Crl \|,\| \Ctj \|\right\}$:
\begin{align}
  \left|(\psi_l^{(n+1)} - \psi_l^{(n)})(z)\right| &\leq |z|\frac{r}{t}\sum_{j=1}^L\left|\left(\tilde\psi^{(n)}_j-\tilde\psi^{(n-1)}_j\right)(z)\right|
  \|\Crl\| \|{\bf T}^{(n)}(z)\| \|\Crj\| \|{\bf T}^{(n-1)}(z)\|
  \\&\leq |z|\Cmax^2\frac{r}{t}\|{\bf T}^{(n)}(z)\| \|{\bf T}^{(n-1)}(z)\|\sum_{j=1}^L\left|\left(\tilde\psi^{(n)}_j-\tilde\psi^{(n-1)}_j\right)(z)\right|.
\end{align}
We now consider $z\in\mathbb{C-R}$. Then ${\bf T}^{(n)}(z)$ and ${\bf T}^{(n-1)}(z)$ have a spectral norm less than $\frac{1}{d(z,\mathbb{R}^+)}$ by item (\ref{pr:it-ineq}) of Proposition \ref{pr:iteration}. Therefore,
\begin{equation}
  \left|(\psi_l^{(n+1)} - \psi_l^{(n)})(z)\right|
  \leq \frac{r \Cmax^2}{t}\frac{|z|}{(d(z,\mathbb{R}^+))^2}\sum_{j=1}^L\left|\left(\tilde\psi^{(n)}_j-\tilde\psi^{(n-1)}_j\right)(z)\right|.
  \label{eq:M1st}
\end{equation}
A similar computation leads to
\begin{equation}
  \left|(\tilde\psi_j^{(n+1)} - \tilde\psi_j^{(n)})(z)\right|
  \leq \Cmax^2\frac{|z|}{(d(z,\mathbb{R}^+))^2}\sum_{l=1}^L\left|\left(\psi^{(n)}_l-\psi^{(n-1)}_l\right)(z)\right|.
  \label{eq:M2nd}
\end{equation}
We now introduce the following maximum:
\[
 M^{(n)}(z)=\max_{l,j}\left\{\Big|(\psi_l^{(n+1)} - \psi_l^{(n)})(z)\Big|,\Big|(\tilde\psi_j^{(n+1)} - \tilde\psi_j^{(n)})(z)\Big|\right\}.
\]
Equations (\ref{eq:M1st}) and (\ref{eq:M2nd}) can then be combined into:
\[
 M^{(n)}(z) \leq \varepsilon(z) M^{(n-1)}(z),
\]
where $\varepsilon(z)=\varepsilon_1\frac{|z|}{(d(z,\mathbb{R}^+))^2}$, with $\varepsilon_1=\max\left\{\frac{r L  \Cmax^2}{t},L  \Cmax^2\right\}$.
We define the following domain: $U=\left\{z \in \mathbb{C}, d(z,\mathbb{R}^+) \geq 2 \varepsilon_1/K, \left|\frac{z}{d(z,\mathbb{R}^+)}\right|\leq 2 \right\}$, with $0\leq K<1$. 
On this domain $U$ we have $M^{(n)}(z) \leq K M^{(n-1)}(z)$.
Hence, for $z \in U$, $\psi_l^{(n)}(z)$ and $\tilde\psi_j^{(n)}(z)$ are Cauchy sequences and, as such, converge.
We denote by $\psi_l(z)$ and $\tilde\psi_j(z)$ their respective limit.

One wants to extend this convergence result on $\mathbb{C}-\mathbb{R}^+$.
We first notice that, as $\psi_l^{(n)}$ is a Stieltj\`es transform whose associated measure has mass $\frac{1}{t}\mathrm{Tr}\,\Crl$, item (\ref{pr:st-bounded}) of Proposition \ref{pr:stieljes} implies
\[
 \psi_l^{(n)}(z)\leq\frac{\frac{1}{t}\mathrm{Tr}{\Crl}}{d(z,\mathbb{R}^+)}.
\]
The $\psi_l^{(n)}$ are thus bounded on any compact set included in $\mathbb{C}-\mathbb{R}^+$, uniformly in $n$.
By Montel's theorem, $\big(\psi_l^{(n)}\big)_{n\in\mathbb{N}}$ is a normal family.
Therefore one can extract a subsequence converging uniformly on compact sets of $\mathbb{C}-\mathbb{R}^+$, whose limit is thus analytic over $\mathbb{C}-\mathbb{R}^+$.
This limit coincides with $\psi_l$ on domain $U$.
The limit of any converging subsequence of $\big(\psi_l^{(n)}\big)$ thus coincides with $\psi_l$ on $U$.
Therefore, these limits all coincide on $\mathbb{C}-\mathbb{R}^+$ with a function analytic on $\mathbb{C}-\mathbb{R}^+$, that we still denote $\psi_l$. 
The converging subsequences of $\big(\psi_l^{(n)}\big)$ have thus the same limit.
We have therefore showed the convergence of the whole sequence $\big(\psi_l^{(n)}\big)_{n\geq0}$ on $\mathbb{C}-\mathbb{R}^+$ towards an analytic function $\psi_l$.
Moreover, as one can check that $\psi_l$ verifies Proposition \ref{pro:stieljes}, we have $\psi_l(z) \in {\cal S}(\mathbb{R}^{+})$.
The same arguments hold for the $\tilde{\psi}_l(z)$.

We have proved the convergence of iterative sequence (\ref{ite-seq-stj}). Taking $z=-\sigma^2$ then yields the convergence of the fixed point algorithm (\ref{eq:point-fixe}). Note that the starting point $(\bs{\delta}^{(0)},\bs{\tilde\delta}^{(0)})$ only needs to verify $\delta_l^{(0)} > 0$,  $\tilde{\delta}_l^{(0)} > 0$ ($l=1,\ldots, L$), as any positive real number can be interpreted as the value at point $z=-\sigma^2$ of some element $s(z) \in \mathcal{S}(\mathbb{R}^+)$.
Moreover, the limits ${\psi}_l(z)$, $\tilde{\psi}_l(z)$ ($l=1,\ldots, L$) of the iterative sequence (\ref{ite-seq-stj}) are positive for any $z=-\sigma^2$ by item (\ref{pr:st-bounded}) of Proposition \ref{pr:stieljes}, as they all are Stieltj\`es transforms.
Therefore, the limits $\drl$, $\dtl$ ($l=1, \ldots, L$) are positive.

\section{A first large system approximation of $\mathbb{E}_\H[\mathrm{Tr}\,\S]$ -- Proof of Proposition \ref{prop:S-appr} }
\label{apx:1st_apprx}

  In this section, if $x$ is a random variable we denote by $\mathring{x}$ the zero mean random variable $\mathring{x}=x-\mathbb{E}(x)$.
  
  We will prove Proposition \ref{prop:S-appr} by deriving the matrix $\bs\Upsilon$ defined by \eqref{eq:propS-appr}, before proving that it satisfies $\frac{1}{t}\mathrm{Tr}\left(\bs\Upsilon \A\right)=\mathcal{O}\left(\frac{1}{t^2}\right)$ for any uniformly bounded matrix $\A$.
  To that end, as the entries of matrices $\Hl$ are Gaussian, we can use the classical Gaussian methods: we introduce here two Gaussian tools,
  an Integration by Parts formula and the Nash-Poincar\'e inequality, both widely used in Random Matrix Theory (see e.g. \cite{pastur2005simple}).
  
  We first present an Integration by Parts formula which provides the expectation of some functionals of Gaussian vectors (see e.g. \cite{novikov1965functionals}).
  \begin{theorem}
   Let $\bs \xi=[\xi_1, \ldots, \xi_M]^T$ a complex Gaussian random vector such that
   $\mathbb{E}[\bs \xi]=\bs 0$,
   $\mathbb{E}[\bs \xi \bs \xi^T]=\bs 0$
   and
   $\mathbb{E}[\bs \xi \bs \xi^H]=\bs \Omega$.
   If $\Gamma : (\bs \xi) \mapsto \Gamma(\bs \xi)$
   is a $\mathcal{C}^1$ complex function polynomially bounded together with its derivatives, then
   \begin{equation}
     \mathbb{E}[\xi_p\Gamma(\bs \xi)]=\sum_{m=1}^M \bs\Omega_{pm} \mathbb{E}\left[\frac{\partial\Gamma(\bs \xi)}{\partial \xi_m^*}\right].
     \label{eq:IPP-th}
   \end{equation}
  \end{theorem}
  In the present context we consider $\bs \xi$ being the vector of the stacked columns of matrices $\Hl$, where the channels $\Hl$ are independent and follow the Kronecker model, i.e.
  $\mathbb{E}_\H\left[\Hk_{ij}\Hlc_{mn}\right]=\delta_{k,l}\frac{1}{t}\Crl_{im}\Ctl_{jn}$.
  Then \eqref{eq:IPP-th} becomes
  \begin{equation}
    \mathbb{E}[\Hl_{ij} \Gamma(\H^{(1)},\ldots, \H^{(L)})]=
    \frac{1}{t} \sum_{m=1}^r\sum_{n=1}^t  \Crl_{im}\Ctl_{jn} \mathbb{E}\left[\frac{\partial\Gamma}{\partial \Hlc_{mn}}\right].
     \label{eq:IPP}
  \end{equation}

  The second useful tool is the Poincar\'e Nash inequality which bounds the variance of certain functionals of Gaussian vectors (see e.g. \cite{pastur2005simple,hachem2008anew}).
  \begin{theorem}
   Let $\bs \xi=[\xi_1, \ldots, \xi_M]^T$ a complex Gaussian random vector such that
   $\mathbb{E}[\bs \xi]=\bs 0$,
   $\mathbb{E}[\bs \xi \bs \xi^T]=\bs 0$
   and
   $\mathbb{E}[\bs \xi \bs \xi^H]=\bs \Omega$.
   If $\Gamma : (\bs \xi) \mapsto \Gamma(\bs \xi)$
   is a $\mathcal{C}^1$ complex function polynomially bounded together with its derivatives, then,
   noting $\nabla_{\bs \xi} \Gamma=[\frac{\partial\Gamma}{\partial \xi_1},\ldots,\frac{\partial\Gamma}{\partial \xi_M}]^T$
   and $\nabla_{\overline{\bs \xi}} \Gamma=[\frac{\partial\Gamma}{\partial \overline\xi_1},\ldots,\frac{\partial\Gamma}{\partial \overline\xi_M}]^T$,
   \begin{equation}
     \var(\Gamma(\bs\xi))\leq
     \mathbb{E}\left[ \nabla_{\bs \xi} \Gamma(\bs \xi)^T \ \bs\Omega \ \overline{\nabla_{\bs \xi} \Gamma(\bs\xi)} \right]
     + \mathbb{E}\left[ \nabla_{\overline{\bs \xi}} \Gamma(\bs \xi)^H \ \bs\Omega \ \nabla_{\overline{\bs \xi}} \Gamma(\bs\xi) \right].
     \label{eq:NPinq-th}
   \end{equation}
  \end{theorem}
  In the following we will use the Nash-Poincar\'e inequality with $\bs \xi$ being the vector of the stacked columns of independent matrices $\Hl$, where the channels $\Hl$ follow the Kronecker model.
  Then \eqref{eq:NPinq-th} becomes
   \begin{equation}
     \var(\Gamma(\H^{(1)},\ldots, \H^{(L)}))\leq
     \frac{1}{t} \sum_{i,m=1}^r \sum_{j,n=1}^t \sum_{l=1}^L \Crl_{im}\Ctl_{jn} 
     \mathbb{E}\left[
     \frac{\partial\Gamma}{\partial \Hl_{ij}} \left(\frac{\partial\Gamma}{\partial \Hl_{mn}}\right)^*
     +\left(\frac{\partial\Gamma}{\partial \Hlc_{ij}}\right)^*  \frac{\partial\Gamma}{\partial \Hlc_{mn}}
     \right].
     \label{eq:NPinq}
   \end{equation}  

  Using these two Gaussian tools we now prove Proposition \ref{prop:S-appr}.
  In order to derive the matrix $\bs\Upsilon$ defined by $\mathbb{E}_\H[\S]= \R + \bs\Upsilon$ we study the entries of $\mathbb{E}_\H[\S]$.
  Using the resolvent identity \eqref{eq:res-id} we have
  $\sigma^2\mathbb{E}_\H[\S_{pq}]=\left[\I-\mathbb{E}_\H[(\S\H\H^H)\right]_{pq}]$.
  We evaluate $\mathbb{E}_\H[(\S\H\H^H)_{pq}]$ by first studying
  $\mathbb{E}_\H \left[ \S_{pi} \Hl_{ij} \H^{(l')}_{qk} \right]$.
  Calculation begins with an integration by parts on $\Hl_{ij}$ \eqref{eq:IPP}:
  \begin{align*}
	  \mathbb{E}_\H \left[ \S_{pi} \Hl_{ij} \H^{(l')*}_{qk} \right]
	  & = \frac{1}{t}\sum_{m,n} \Crl_{im}\Ctl_{jn} \mathbb{E}_\H\left[\frac{\partial(\S_{pi}\H^{(l')*}_{qk})}{\partial\Hlc_{mn}}\right]
	  \\
	  & = \frac{1}{t} \sum_{m,n}  \Crl_{im}\Ctl_{jn} \mathbb{E}_\H\left[\S_{pi}\delta_{l,l'}\delta_{q,m}\delta_{k,n}+\H^{(l')*}_{qk}\frac{\partial \S_{pi}}{\partial \Hlc_{mn}}\right].
  \end{align*}
  As $\frac{\partial \S_{pi}}{\partial \Hlc_{mn}}=-\Big( \S\frac{\partial \S^{-1}}{\partial \Hlc_{mn}}\S \Big)_{pi}=-(\S\H)_{pn} \S_{mi}$, we obtain
  \begin{align*}
    \mathbb{E}_\H\left[\S_{pi}\Hl_{ij}\H^{(l')*}_{qk}\right]
    &= \frac{1}{t}\Crl_{iq}\Ctl_{jk} \mathbb{E}_\H[\S_{pi}]\delta_{l,l'}
      -\frac{1}{t}\sum_{m,n} \Crl_{im} \Ctl_{jn} \mathbb{E}_\H\left[\H^{(l')*}_{qk} (\S \H)_{pn} \S_{mi}\right]
	  \\
	  & = \frac{1}{t}\Crl_{iq} \Ctl_{jk}\mathbb{E}_\H[\S_{pi}]\delta_{l,l'} - \frac{1}{t} \sum_n \Ctl_{jn} \mathbb{E}_\H\left[\H^{(l')*}_{qk}(\S\H)_{pn} (\Crl\S)_{ii}\right].
  \end{align*}
  Summing over $i$, $l$ and $l'$ then leads to:
  \[
	  \mathbb{E}_\H\left[(\S\H)_{pj}\H_{qk}^*\right]
	  = \sum_l\frac{1}{t}\mathbb{E}_\H[(\S\Crl)_{pq}]\Ctl_{jk} -\sum_{n,l}{\Ctl_{jn}\mathbb{E}_\H\left[\H_{qk}^*(\S\H)_{pn} \frac{1}{t}\Tr(\S\Crl) \right]}.
  \]
  To separate the terms under the last expectation, we denote
  $\eta_l=\frac{1}{t}\Tr(\S\Crl)=\alpha_l+\mathring{\eta_l}$,
  where $\alpha_l=\mathbb{E}_\H[\eta_l]$.
  We can then write
  $\mathbb{E}_\H\left[\H_{qk}^*(\S\H)_{pn} \eta_l) \right]=\alpha_l\mathbb{E}_\H\left[\H_{qk}^*(\S\H)_{pn} \right]+\mathbb{E}_\H\left[\H_{qk}^*(\S\H)_{pn} \mathring{\eta_l} \right]$, hence
  \begin{equation}
	  \mathbb{E}_\H\left[(\S\H)_{pj}\H_{qk}^*\right]
	  = \sum_{l}\frac{1}{t}\mathbb{E}_\H[(\S\Crl)_{pq}]\Ctl_{jk}  -\sum_{n,l}\alpha_l\Ctl_{jn}\mathbb{E}_\H\left[(\S\H)_{pn}\H_{qk}^*\right] - \bs\Xi^{(p,q)}_{jk},
	  \label{eq:bef-Delta}
  \end{equation}
  where
  $\bs\Xi^{(p,q)}_{jk}=\sum_{n}{\mathbb{E}_\H\left[\H_{qk}^*(\S\H)_{pn}\sum_l \mathring{\eta_l}\Ctl_{jn}\right]}$.
  We here notice the presence of
  $\mathbb{E}_\H\left[(\S\H)_{p(*)}\H_{qk}^*\right]$
  on both sides of equality \eqref{eq:bef-Delta}.
  Hence, let us denote
  $\Delta^{(p,q)}_{jk}=\mathbb{E}_\H\left[(\S\H)_{pj}\H_{qk}^*\right]$.
  Then \eqref{eq:bef-Delta} becomes
  \[
	  \Delta^{(p,q)}_{jk}
	  = \sum_{l}\frac{1}{t}\mathbb{E}_\H[(\S\Crl)_{pq}]\Ctl_{jk}  -\Big(\sum_l\alpha_l\Ctl\Delta^{(p,q)}\Big)_{jk} - \bs\Xi^{(p,q)}_{jk}.
  \]
  Recalling that $\tilde\R=\left(\sigma^2\left(\I_t+\sum_l\alpha_l\Ctl\right)\right)^{-1}$, this leads to
  \[
	  \Delta^{(p,q)} =\sigma^2\sum_l\frac{1}{t}\mathbb{E}_\H[(\S\Crl)_{pq}]\tilde\R \Ctl - \sigma^2\tilde\R\bs\Xi^{(p,q)}.
  \]
  We now come back to the calculation of $\mathbb{E}_\H[\S_{pq}]=\frac{1}{\sigma^2}(\I_r-\mathbb{E}_\H[(\S\H\H^H)])_{pq}$ by noticing that
  $\mathbb{E}_\H[(\S\H\H^H)_{pq}]=\sum_j\mathbb{E}_\H\left[(\S\H)_{pj}\H_{qj}^*\right]=\Tr(\Delta^{(p,q)})$.
  Therefore
  \[
	  \mathbb{E}_\H[\S_{pq}]
	   =\frac{\delta_{p,q}}{\sigma^2}-\sum_l \tilde\alpha_l \mathbb{E}_\H[(\S\Crl)_{pq}] +\Tr\left(\tilde\R\bs\Xi^{(p,q)}\right),
  \]
  as $\tilde\alpha_l=\frac{1}{t}\Tr\left(\tilde\R\Ctl\right)$ \eqref{eq:def-alphas}.
  Coming back to the definition of matrix $\bs\Xi^{(p,q)}$, we notice that
  $
   \Tr\left(\tilde\R\bs\Xi^{(p,q)}\right)
    = \sum_l\mathbb{E}_\H\left[ \mathring{\eta}_l (\S \H \Ct^{(l)T} \Rt^T \H^H )_{pq} \right]
  $.
  Hence the matrix $\mathbb{E}_\H[\S]$ can be written as
  \[
	  \mathbb{E}_\H[\S]
	    =\frac{1}{\sigma^2} \I_r - \mathbb{E}_\H[\S] \sum_l \tilde\alpha_l \Crl + \sum_l\mathbb{E}_\H\left[ \mathring{\eta}_l \S \H \Ct^{(l)T} \Rt^T \H^H  \right].
  \]
  And finally,
  \begin{equation}
	  \mathbb{E}_\H[\S]=\R + \bs\Upsilon,
	  \label{eq:S-eqv}
  \end{equation}
  where $\R=\left(\sigma^2\left(\I_t+\sum_l\tilde\alpha_l\Crl\right)\right)^{-1}$ and where the matrix $\bs\Upsilon$ is defined as
  \begin{equation}
   \bs\Upsilon=\sigma^2\sum_l\mathbb{E}_\H\left[ \mathring{\eta}_l \S \H \Ct^{(l)T} \Rt^T \H^H  \right] \R .
   \label{eq:Ups-def}
  \end{equation}
  
  To end the proof of Proposition \ref{prop:S-appr} we now need to prove that
  $\frac{1}{t}\mathrm{Tr}\left(\bs\Upsilon \A\right)=\mathcal{O}\left(\frac{1}{t^2}\right)$ for any uniformly bounded matrix $\A$.
  Let $\A$ be a $r \times r$ matrix uniformly bounded in $r$.
  Using \eqref{eq:Ups-def},
  \begin{align*}
   \frac{1}{t}\mathrm{Tr}\left(\bs\Upsilon \A\right) 
   &= \frac{\sigma^2}{t}\sum_l\mathbb{E}_\H\left[\mathring{\eta_l}\mathrm{Tr} \left(\S\H\Ct^{(l)T}\tilde\R^T\H^H\R\A  \right) \right] 
   \\
   &= \frac{\sigma^2}{t}\sum_l\mathbb{E}_\H\Bigg[\mathring{\eta_l} \ \overset{\circ }{ \overparenthesis{\mathrm{Tr}(\S\H\Ct^{(l)T}\tilde\R^T\H^H\R\A)} } \Bigg].
  \end{align*}
  We can now bound $\frac{1}{t}\mathrm{Tr}\left(\bs\Upsilon \A\right)$ thanks to Cauchy-Schwartz inequality.
  \begin{align}
   \left|\frac{1}{t}\mathrm{Tr}\left(\bs\Upsilon \A\right)\right|
   \leq \frac{\sigma^2}{t}&\sum_l
      \sqrt{\mathbb{E}_\H \left[ \left| \mathring \eta_l \right|^2 \right]} \ 
      \sqrt{\mathbb{E}_\H \left[\Bigg|\overset{\circ }{ \overparenthesis{ \mathrm{Tr} \left(\S\H\Ct^{(l)T}\tilde\R^T\H^H\R\A  \right)}} \Bigg|^2\right]} \notag
   \\
   = &\frac{\sigma^2}{t}\sum_l\sqrt{\var\left(\eta_l\right)}\sqrt{\var\left(\mathrm{Tr} \left(\S\H\Ct^{(l)T}\tilde\R^T\H^H\R\A  \right)\right)},
   \label{eq:CS-2var}
  \end{align}
  as $\mathbb{E}_\H\left[ \left| \mathring{x} \right|^2 \right]=\var\left(x\right)$ for any random variable $x$.
  We first prove that $\var\left(\eta_l\right)= \mathcal{O}\left(\frac{1}{t^2}\right)$.
  The Nash-Poincar\'e inequality \eqref{eq:NPinq} states that
  \begin{equation}
   \var(\eta_l) \leq
     \frac{1}{t} \sum_{i,j,m,n,k} \Crk_{im}\Ctk_{jn} 
     \mathbb{E}\left[
     \frac{\partial\eta_l}{\partial\Hk_{ij}} \bigg( \frac{\partial\eta_l}{\partial\Hk_{mn}} \bigg)^*
     +
     \bigg( \frac{\partial\eta_l}{\partial\Hkc_{ij}} \bigg)^* \frac{\partial\eta_l}{\partial\Hkc_{mn}} 
     \right].
     \label{eq:NP_etal}
  \end{equation}
  As $\frac{\partial \S_{pq}}{\partial \Hk_{ij}}
    =-\big(  \S\frac{\partial \S^{-1}}{\partial \Hk_{ij}}\S  \big)_{pq}
    =-\S_{pi}(\H^H\S)_{jq}$
  we can derive $\frac{\partial\eta_l}{\partial\Hk_{ij}}$:
  \[
   \frac{\partial\eta_l}{\partial\Hk_{ij}}
     =\frac{1}{t}\mathrm{Tr}\bigg(\frac{ \partial \S }{ \partial \Hk_{ij} } \Crl \bigg)
     = \frac{1}{t}\sum_{p,q}\frac{ \partial \S_{pq} }{ \partial \Hk_{ij} } \Crl_{qp}
     = -\frac{1}{t} (\H^H\S \Crl \S )_{ji}.
  \]
  Similarly we obtain
    $\frac{\partial\eta_l}{\partial\Hkc_{ij}} = -\frac{1}{t} (\S \Crl \S\H)_{ij}$.
  Therefore \eqref{eq:NP_etal} becomes
  \begin{align*}
      \var(\eta_l) \leq
     \frac{1}{t^3} \sum_{i,j,m,n,k}  \Crk_{im}\Ctk_{jn} \mathbb{E} & \left[ 
       (\H^H\S \Crl \S )_{ji} (\H^H\S \Crl \S )_{nm}^*
       +
       (\S \Crl \S\H)_{ij}^*(\S \Crl \S\H)_{mn}
       \right]
     \\
     = \frac{1}{t^3}  \sum_k \mathbb{E} & \left[
     \mathrm{Tr} \left( (\H^H\S \Crl \S) \Crk (\H^H\S \Crl \S )^H \Ct^{(k)T} \right) \right.
     \\
     & \left.+\mathrm{Tr}\left(\Ct^{(k)T} (\S \Crl \S\H)^H\Crk (\S \Crl \S\H)\right)
     \right].
  \end{align*}
  Then, using the inequality $|\mathrm{Tr}(\B_1\B_2)|\leq\|\B_1\|\mathrm{Tr}\,\B_2$, where $\B_2$ is non-negative hermitian, for both traces in the above expression,
  \begin{align}
      \var(\eta_l)
      &\leq \frac{2}{t^3} \|\Crl\|^2 \sum_k \|\Crk\| \ \mathbb{E} \left[ \|\S\|^4 \mathrm{Tr} \left( \H \Ct^{(k)T} \H^H \right) \right]
      \notag
      \\
      &\leq \frac{2}{t^3} \|\Crl\|^2 \sum_k \|\Crk\| \|\Ctk\| \ \mathbb{E} \left[ \|\S\|^4 \mathrm{Tr} \left( \H \H^H \right) \right]
      \notag
      \\
      &\leq \frac{1}{t^2} \frac{2 L C_{\sup}^4}{\sigma^8} \mathbb{E} \left[  \frac{1}{t}\mathrm{Tr} \left( \H \H^H \right) \right],
      \label{eq:var_etal_final}
  \end{align}
   where the second inequality follows from $\| \S \|\leq\frac{1}{\sigma^2}$ and from the definition of $C_{\sup}$:
   \begin{equation}
    \Csup=\sup_t \Cmax= \sup_t \ \left\{\max_{k,l}\left\{\| \Crk \|,\| \Ctl \|\right\}\right\}.
    \label{eq:Csup}
   \end{equation}
   The hypotheses of Proposition \ref{prop:S-appr} ensure that $\Csup<+\infty$.
   We now prove that $\mathbb{E}\left[\frac{1}{t}\mathrm{Tr} \left( \H \H^H \right)\right]=\mathcal{O}\left(1\right)$.
   Using the fact that the channels $\Hl$ are independent and follow the Kronecker model, that is
   $\mathbb{E}_\H\left[\Hk_{ij}\Hlc_{mn}\right]=\delta_{k,l}\frac{1}{t}\Crl_{im}\Ctl_{jn}$,
  \begin{align*}
    \mathbb{E}_\H\left[\frac{1}{t}\mathrm{Tr}\left(\H \H^H \right)\right]
    &= \frac{1}{t} \sum_{i,j,k,l} \mathbb{E}_\H \left[ \Hk_{ij} \Hlc_{ij}\right]
    = \frac{1}{t^2} \sum_{i,j,l} \Crl_{ii} \Ctl_{jj} 
    = \frac{1}{t^2} \sum_{l} \mathrm{Tr}\,\Crl \mathrm{Tr}\, \Ctl
    \\
    &\leq \frac{r}{t} \sum_{l} \|\Crl\| \|\Ctl\|  
    \leq \frac{r}{t} L \Csup^2.
   \end{align*}
   Therefore we proved that $\mathbb{E}_\H\left[\frac{1}{t}\mathrm{Tr} \left( \H \H^H \right)\right]=\mathcal{O}\left(1\right)$.
   Coming back to \eqref{eq:var_etal_final} gives
   $\var(\eta_l)\leq \frac{1}{t^2} \left(\frac{r}{t}\frac{2 \Csup^6 L^2}{\sigma^8}  \right)$,
   hence
   $\var(\eta_l)=\mathcal{O}\left(\frac{1}{t^2}\right)$.
  
  We evaluate similarly the behavior of the second term of the right-hand side of \eqref{eq:CS-2var} and we obtain $\var\left(\mathrm{Tr} \left(\S\H\Ct^{(l)T}\tilde\R^T\H^H\R\A  \right)\right)\leq \frac{k}{\sigma^{12}}\left(1+\frac{1}{\sigma^2}\right) \| \A \|^2 =\mathcal{O}\left(1\right)$, where $k$ does not depend on $\sigma^2$ nor on $t$.
  As $\var (\eta_l)  = \mathcal{O}\left(\frac{1}{t^2}\right)$, we eventually have
  \[
   \frac{1}{t}\mathrm{Tr}(\bs\Upsilon \A) = \mathcal{O}\left(\frac{1}{t^2}\right),
  \]
  which completes the proof of Proposition \ref{prop:S-appr}.
  \begin{remark}
    \label{rem:intgbl_trGA}
     Note that, as
     $\var(\eta_l)\leq \frac{1}{\sigma^8 t^2} \left(2\frac{r}{t}\Csup^6 L^2  \right)$
     and
     $\var\left(\mathrm{Tr} \left(\S\H\Ct^{(l)T}\tilde\R^T\H^H\R\A  \right)\right)\leq \frac{1}{\sigma^{12}} \left(k \| \A \|^2\left(1+\frac{1}{\sigma^2}\right)\right)$,
     \eqref{eq:CS-2var} leads to
     $\frac{1}{t}\mathrm{Tr}(\bs\Upsilon \A) \leq \frac{1}{\sigma^8 t^2}P\left(\frac{1}{\sigma^2}\right)$,
     where $P$ is a polynomial with real positive coefficients which do not depend on $\sigma^2$ nor on $t$.
  \end{remark}  

\section{A refined large system approximation of $\mathbb{E}_\H[\mathrm{Tr}\,\S]$ -- Proof of Proposition \ref{prop:2nd-appr}}
\label{apx:alpha-delta}

  We prove in this section that
  $\frac{1}{t}\mathrm{Tr}(\R\A)=\frac{1}{t}\mathrm{Tr}(\T\A)+\mathcal{O}\left(\frac{1}{t^2}\right)$
  for any $r \times r$ matrix $\A$ uniformly bounded in r.
  We first note that the difference $\frac{1}{t} \mathrm{Tr}\left( \R \A \right) -\frac{1}{t} \mathrm{Tr}\left( \T\A \right)$ can be written as
  \begin{equation}
   \label{eq:tr_R-T}
   \frac{1}{t} \mathrm{Tr}\left((\R-\T)\A\right) 
   =\frac{1}{t} \mathrm{Tr}\left(\R\left(\T^{-1}-\R^{-1}\right)\T\A\right) 
   =-\frac{\sigma^2}{t} \sum_l(\tilde\alpha_l-\tilde\delta_l)\mathrm{Tr}\left(\R\Crl\T\A\right).
  \end{equation}
  As $\| \T \| \leq \frac{1}{\sigma^2}$ and $\| \R \| \leq \frac{1}{\sigma^2}$, expression \eqref{eq:tr_R-T} yields
   \begin{equation}
      \label{ineq:tr_R-T}
      \frac{1}{t} \left| \mathrm{Tr}\left((\R-\T)\A\right)  \right|
	\leq \frac{r}{t} \frac{\Csup\|\A\|}{\sigma^2} \sum_l \big|\tilde\alpha_l-\tilde\delta_l \big|,
   \end{equation}
  where $\Csup<+\infty$ is defined by \eqref{eq:Csup}.
  We now consider the difference $\frac{1}{t}\mathrm{Tr}(\Rt \tilde\A)-\frac{1}{t}\mathrm{Tr}(\Tt \tilde\A)$ for any $t \times t$ matrix $\tilde\A$ uniformly bounded in t,
  which can be derived similarly:
  \begin{equation}
      \frac{1}{t}\left|\mathrm{Tr}\left( \big(\Rt-\Tt \big)\tilde\A\right)\right|
      \leq \frac{\Csup \| \tilde\A \|}{\sigma^2} \sum_l \left| \alpha_l-\delta_l \right|.
      \label{ineq:tr_Rt-Tt}
  \end{equation}
  Taking
  $\A=\Crk$ in \eqref{ineq:tr_R-T},
  $\tilde\A=\Ctk$ in \eqref{ineq:tr_Rt-Tt}
  and using Proposition \ref{prop:S-appr}
  gives
  \begin{align}
     &\big| \alpha_k - \delta_k \big| \leq \frac{r}{t} \frac{\Csup^2}{\sigma^2} \sum_l \big|\tilde\alpha_l-\tilde\delta_l \big| + \mathcal{O} \left( \frac{1}{t^2} \right),
     \\
     &\big| \tilde\alpha_k - \tilde\delta_k \big| \leq \frac{\Csup^2}{\sigma^2}   \sum_l \left| \alpha_l-\delta_l \right|, 
      \label{eq:alpha_t-delta_t}
  \end{align}
  which leads to
  \[
     \left( 1 - \frac{r}{t} \frac{\Csup^4 L^2}{\sigma^4} \right) \sum_k \big| \alpha_k - \delta_k \big|
      \leq \mathcal{O} \left( \frac{1}{t^2} \right).
  \]
  Therefore it is clear that there exists $\sigma_0^2$ such that
  $\big| \alpha_k - \delta_k \big| =\mathcal{O} \left( \frac{1}{t^2} \right)$ for $\sigma^2>\sigma_0^2$
  for any $k \in \{1, \ldots, L\}$.
  In particular,
  $\big| \alpha_k - \delta_k \big| \xrightarrow{t\to \infty}~0$ for $\sigma^2>\sigma_0^2$.
  We now extend this result to any $\sigma^2>0$.
  To this end, similarly to Appendix \ref{apx:delta_exist-unique}, it is useful to consider
  $\alpha_l$ and $\delta_l$ as functions of the parameter $(-\sigma^2) \in \mathbb{R}^-$
  and to extend their domain of validity from $\mathbb{R}^-$ to $\mathbb{C}-\mathbb{R}^+$ in order to use the results about Stieltj\`es transforms.
  The function $\delta_l(z)$ then corresponds to the function $\psi_l(z)$ of Appendix \ref{apx:delta_exist-unique} and therefore belongs to $\mathcal{S}(\mathbb{R}^+)$ with an associated measure of mass $\frac{1}{t}\mathrm{Tr}\,\Crl$, for $l = 1,\ldots, L $.
  It is easy to check that function
  $\alpha_l(z)$ also belongs to $\mathcal{S}(\mathbb{R}^+)$ with an associated measure of mass $\frac{1}{t}\mathrm{Tr}\,\Crl$ for any $l \in \left\{ 1,\ldots, L \right\}$.
  Hence, by Proposition \ref{pr:stieljes} (\ref{pr:st-bounded}), we can upper bound the Stieltj\`es transforms $\alpha_l(z)$ and $\delta_l(z)$ on $\mathbb{C-R}^+$, yielding:
  \[
   \left| \alpha_l(z) - \delta_l(z) \right|
   \leq 2 \frac{\frac{1}{t}\mathrm{Tr}\,\Crl}{d(z,\mathbb{R}^+)}
   \leq 2 \frac{\frac{r}{t} \Csup}{d(z,\mathbb{R}^+)}.
  \]
  The
  $(\alpha_l(z)-\delta_l(z))_{t\in\mathbb{N}}$
  are thus bounded on any compact set included in $\mathbb{C}-\mathbb{R}^+$, uniformly in $t$.
  Moreover $(\alpha_l(z)-\delta_l(z))_{t\in\mathbb{N}}$ is a family of analytic functions.
  Using Montel's theorem similarly to Appendix \ref{apx:delta_exist-unique}, we obtain that
  $\big| \alpha_l(z) - \delta_l(z) \big| \xrightarrow{t\to \infty}~0$ on $\mathbb{C-R}^+$ for any $l \in \{1,\ldots,L\}$, thus in particular
  \begin{equation}
    \big| \alpha_l - \delta_l \big| \xrightarrow{t\to \infty}~0
   \label{eq:alpha->delta}
  \end{equation}
  for any $\sigma^2>0$, $l \in \{1,\ldots,L\}$,
  which, used in \eqref{eq:alpha_t-delta_t}, yields
  \begin{equation}
    \big| \tilde\alpha_l - \tilde\delta_l \big| \xrightarrow{t\to \infty}~0
    \label{eq:alpha_t->delta_t}
  \end{equation}
  for any $\sigma^2>0$, $l \in \{1,\ldots,L\}$.
  Using \eqref{eq:alpha_t->delta_t} in \eqref{ineq:tr_R-T} and \eqref{eq:alpha->delta} in \eqref{ineq:tr_Rt-Tt} gives
  \begin{align}
    &\frac{1}{t}\mathrm{Tr\left(\A (\R-\T) \right)} \xrightarrow{t\to \infty} 0,
    \label{eq:trR->trT}
    \\
    &\frac{1}{t}\mathrm{Tr\left(\tilde\A (\Rt-\Tt) \right)} \xrightarrow{t\to \infty} 0.
    \label{eq:trRt->trTt}
  \end{align}

  We now refine \eqref{eq:trR->trT} and \eqref{eq:trRt->trTt} to prove that these two traces are $\mathcal{O}\left(\frac{1}{t^2}\right)$.
  Taking $\A=\Crl$ in \eqref{eq:tr_R-T} leads to
  $\alpha_k-\delta_k = -\frac{\sigma^2}{t} \sum_l(\tilde\alpha_l-\tilde\delta_l)\mathrm{Tr}\left(\Crl\T\Crk\R\right) + \frac{1}{t}\mathrm{Tr}\left( \Crk \bs\Upsilon \right)$,
  where $\bs\Upsilon=\mathbb{E}_\H[\S]-\R$,
  and similarly
  $\tilde\alpha_k-\tilde\delta_k = -\frac{\sigma^2}{t} \sum_l(\alpha_l-\delta_l)\mathrm{Tr}\left(\Ctl\tilde\T\Ctk\tilde\R\right)$.
  We can rewrite these two equalities under the following matrix form:
  \begin{equation}
    \left( \I_{2L} - \N(\R,\T,\tilde\R,\tilde\T) \right)
    \begin{bmatrix}  \bs\alpha -\bs\delta \\ \bs{\tilde\delta}-\bs{\tilde\alpha} \end{bmatrix}
    =
    \begin{bmatrix}  \bs\varepsilon \\ \bs{0} \end{bmatrix},
    \label{eq:delta-alpha_matrix}
  \end{equation}
  where $\bs\varepsilon$ is a $L \times 1$ vector whose entries defined by
  $\bs\varepsilon_k = \frac{1}{t}\mathrm{Tr}\left( \Crk \bs\Upsilon \right)$
  verify  $\bs\varepsilon_k = \mathcal{O} \left(\frac{1}{t^2}\right)$, $k= 1, \ldots, L$, by Proposition \ref{prop:S-appr},
  and where matrix $\N(\R,\T,\tilde\R,\tilde\T)$ is defined by
  \begin{equation}
    \N(\R,\T,\tilde\R,\tilde\T)=\sigma^2\begin{bmatrix} \bs 0 & \B(\R,\T) \\  \tilde \B(\tilde \R,\tilde \T ) & \bs 0  \end{bmatrix},
    \label{eq:N_defin}
  \end{equation}
  where matrices $\B(\R,\T)$ and $\tilde\B(\tilde\R,\tilde\T)$ are $L \times L$ matrices whose entries are defined by
  $\B_{kl}(\R,\T)=\frac{1}{t}\mathrm{Tr}\left( \Crl\T\Crk \R\right)$
  and
  $\tilde\B_{kl}(\tilde\R, \tilde \T)=\frac{1}{t}\mathrm{Tr} \big( \Ctl\tilde\T\Ctk\tilde\R \big)$.
  Besides, taking $\A=\Crl\T\Crk$ in \eqref{eq:trR->trT} and $\tilde\A=\Ctl\tilde\T\Ctk$ in \eqref{eq:trRt->trTt} leads to
  \begin{equation}
   \left\{ \begin{array}{l}
      \frac{1}{t}\mathrm{Tr}\left(\Crl\T\Crk\R \right) \xrightarrow{t\to \infty} \frac{1}{t}\mathrm{Tr\left(\Crl\T\Crk\T \right)},
      \\
      \frac{1}{t}\mathrm{Tr}\left(\Ctl\tilde\T\Ctk\tilde\R \right) \xrightarrow{t\to \infty} \frac{1}{t}\mathrm{Tr\left(\Ctl\tilde\T\Ctk\tilde\T \right)}.
    \end{array}\right.
      \label{eq:tr_CRCT}
  \end{equation}
  Hence
  $\B_{kl}(\R,\T) \xrightarrow{t\to \infty} \A_{kl}(\T)$
  and 
  $\tilde\B_{kl}(\tilde\R,\tilde\T) \xrightarrow{t\to \infty} \tilde\A_{kl}(\tilde\T)$,
  where matrices
  $\A(\T)$ and $\tilde\A(\tilde\T)$ are defined by \eqref{eq:A-Atilde}.
  We now introduce the following lemma:

  \begin{lemma}
  \label{lm:rsp}
  Let $\T$, $\tilde\T$ be the matrices defined by (\ref{eq:canoniqueter}) with $(\bs{\delta},\bs{\tilde\delta})$ verifying the canonical equation (\ref{eq:canonique}) with $\Q=\I_t$).
  Let $\A(\T)$ and $\tilde\A(\T)$ be the $L \times L$ matrices whose entries are defined by
  $\A_{kl}({\bf T})=\frac{1}{t}\mathrm{Tr}\left(\Crk\T\Crl\T\right)$ and
  $\tilde\A_{kl}(\tilde\T)=\frac{1}{t}\mathrm{Tr}(\Ctk\tilde\T\Ctl\tilde\T)$
  and $\M(\T,\tilde\T)$ the matrix defined by
  \[
   \M(\T,\tilde\T)= \sigma^2\begin{bmatrix} 0 & \A(\T) \\ \tilde\A(\tilde\T) & 0 \end{bmatrix}.
  \]
  Assume that, for every $l \in \left\{ 1, \ldots, L\right\}$, $\sup_t\|\Crl\|<+\infty$, $\sup_t\|\Ctl\|<+\infty$,
  $\inf_t\left(\frac{1}{t}\mathrm{Tr\,} \Crl\right)>0$ and $\inf_t\left(\frac{1}{t}\mathrm{Tr}\,\Ctl \right)>0$.
  Then there exists $k_0>0$ and $k_1<\infty$ both independent of $\sigma^2$ such that
  \begin{enumerate}[(\it i)]
    \item \label{it:rspM}$ \sup_t\left[ \rho\left(\M)\right) \right]
      \leq 1 - \frac{k_0 \sigma^2}{(\sigma^2+k_1)^2} <1$,
    \item \label{it:rspAAt} $\sup_t\left[ \rho\left(\sigma^4\tilde\A(\tilde\T)\A(\T)\right) \right]
      \leq \left( 1 - \frac{k_0 \sigma^2}{(\sigma^2+k_1)^2} \right)^2 <1$,
    \item \label{it:tnorm} $\sup_t\Big[ \tnorm[(\I_{2L}-\M(\T,\tilde\T))^{-1}] \Big]
      \leq \frac{(\sigma^2+k_1)^2}{k_0 \sigma^2}$,
  \end{enumerate}
  where $\tnorm[\cdot]$ is the max-row $\ell_1$ norm defined by
  $\displaystyle\tnorm[\P]=\max_{j\in\left\{ 1,\ldots, M \right\}} \sum_{k=1}^N \left| \P_{jk} \right|$ for a $M \times N$ matrix $\P$.
  \end{lemma} \medskip
  
  \begin{proof}
  Using the expression of $\T^{-1}=\sigma^2 (\I_r + \sum_k \tilde\delta_k \Crk)$, $\delta_l$ can be written as:
  \begin{align*}
  \delta_l
  &=\frac{1}{t}\mathrm{Tr}(\Crl{\bf T}{\bf T}^{-1}{\bf T})
  \\
  &=\frac{\sigma^2}{t}\mathrm{Tr}(\Crl{\bf T}{\bf T})+\frac{\sigma^2}{t}\sum_{k=1}^L\tilde{\delta_k}\mathrm{Tr}(\Crl{\bf T}\Crk{\bf T}).
  \end{align*}
  Similarly it holds that
  $\tilde{\delta}_l=\frac{\sigma^2}{t}\mathrm{Tr}(\Ctl\tilde{\bf T}\tilde{\bf T})+\frac{\sigma^2}{t}\sum_{k=1}^L{\delta_k}\mathrm{Tr}(\Ctl\tilde{\bf T}\Ctk\tilde{\bf T})$.
  Thus,
  \[
  \begin{bmatrix} {\bs \delta} \\ \tilde{\bs \delta} \end{bmatrix}
  = \sigma^2 \begin{bmatrix} 0 & {\bf A}({\bf T}) \\ \tilde{\bf A}(\tilde{\bf T}) & 0 \end{bmatrix}
  \begin{bmatrix} {\bs \delta} \\ \tilde{\bs \delta} \end{bmatrix}
  + \begin{bmatrix} \w \\ \tilde\w \end{bmatrix},
  \]
  where $\w$ and $\tilde\w$ are $L \times 1$ vectors such that
  $\w_l=\frac{\sigma^2}{t}\mathrm{Tr}(\Crl{\bf T}{\bf T})$
  and
  $\tilde\w_l=\frac{\sigma^2}{t}\mathrm{Tr}(\Ctl \tilde\T \tilde\T)$.
  This equality is of the form ${\bf u}=\M(\T,\tilde\T) {\bf u}+{\bf v}$, with
  ${\bf u}=\big[ \bs\delta, \bs{\tilde\delta} \big]^T$
  and
  ${\bf v}=\big[ \w, \tilde\w \big]^T$,
  the entries of ${\bf u}$ and ${\bf v}$ being positive, and the entries of $\M(\T,\tilde\T)$ non-negative.
  A direct application of Corollary 8.1.29 of \cite{horn1990matrix} then implies
  $\rho(\M(\T,\tilde\T)) \leq 1-\frac{\min v_l}{\max u_l}$.
  
  We first briefly consider $\sup_t\left\{\max u_l\right\}$.
  As $\|\T\|\leq\frac{1}{\sigma^2}$ and $\|\Crl\| \leq \Csup$ we have
  \begin{equation}
   \delta_l=\frac{1}{t}\mathrm{Tr}\left( \Crl\T \right)\leq \frac{r}{\sigma^2 t} \Csup.
  \end{equation}
  Similarly, as $\|\tilde\T\|\leq\frac{1}{\sigma^2}$ and $\|\Ctl\| \leq \Csup$,
  \begin{equation}
    \tilde\delta_l =\frac{1}{t}\mathrm{Tr}\left( \Ctl\tilde\T \right)\leq \frac{1}{\sigma^2} \Csup.
    \label{eq:deltal_bound}
  \end{equation}
  As $t/r \xrightarrow{t\to\infty}c>0$ we have that $\sup_t\left[ r/t \right]< +\infty$.
  Therefore $\sup_t\left\{\max u_l\right\}\leq \frac{\lambda_0}{\sigma^2} <\infty$, where
  $\lambda_0= \Csup \max\left\{1, \sup_t\left[ r/t \right] \right\}$.
  
  We now consider
  $\inf_t\left\{\min v_l\right\}=\inf_t\left\{ \min_{k,l} \left\{ \frac{\sigma^2}{t}\mathrm{Tr}(\Crl\T\T), \frac{\sigma^2}{t}\mathrm{Tr}(\Ctk\tilde\T\tilde\T) \right\}\right\}$.
  We will use the Cauchy-Schwarz inequality:
  \begin{equation}
   \left|\mathrm{Tr}(\A\B)\right| \leq \sqrt{\mathrm{Tr}(\A\A^H)} \ \sqrt{\mathrm{Tr}(\B\B^H)}.
   \label{eq:trace-inq}
  \end{equation}
  Taking $\A=\left(\Crl\right)^{1/2}\T$ and $\B=\left(\Crl\right)^{1/2}$ in \eqref{eq:trace-inq} leads to
  \begin{equation}
   \frac{1}{t}\mathrm{Tr}\left( \Crl\T\T \right)
      \geq \frac{\left(\frac{1}{t}\mathrm{Tr}\left( \Crl \T \right)\right)^2}{\frac{1}{t}\mathrm{Tr}\, \Crl }
      = \frac{\delta_l^2}{\frac{1}{t}\mathrm{Tr}\, \Crl }.
    \label{eq:vl_min}
  \end{equation}
  We use again inequality \eqref{eq:trace-inq}, this time with $\A=\left(\Crl\right)^{1/2} \T^{1/2}$ and $\B=\T^{-1/2}\left(\Crl\right)^{1/2}$. Then,
  \begin{equation}
   \delta_l=\frac{1}{t}\mathrm{Tr}\left( \Crl\T\right)
      \geq \frac{\left(\frac{1}{t}\mathrm{Tr} \, \Crl \right)^2}{\frac{1}{t}\mathrm{Tr}\left( \Crl \T^{-1} \right)}.
      \label{eq:deltal_min}
  \end{equation}
  Thanks to \eqref{eq:deltal_bound}, $\|\T^{-1}\| = \| \sigma^2 (\I_r + \sum_l \tilde\delta_l \Crl) \| \leq \sigma^2 + L \Csup^2$.
  Hence \eqref{eq:deltal_min} leads to 
  \begin{equation}
   \delta_l
      \geq \frac{\frac{1}{t}\mathrm{Tr} \, \Crl }{\|\T^{-1}\|}
      \geq \frac{\frac{1}{t}\mathrm{Tr} \, \Crl }{\sigma^2 + L \Csup^2}.
    \label{eq:deltal_min_fin}
  \end{equation}
  Eventually, using \eqref{eq:deltal_min_fin} in \eqref{eq:vl_min} gives
  \begin{equation}
   \frac{1}{t}\mathrm{Tr}\left( \Crl\T\T \right)
      \geq \frac{\frac{1}{t}\mathrm{Tr} \, \Crl }{\left(\sigma^2 + L \Csup^2\right)^2}.
  \end{equation}
  Similarly, we prove that
  \[
     \frac{1}{t}\mathrm{Tr}\left( \Ctl\tilde\T\tilde\T \right)
      \geq \frac{\frac{1}{t}\mathrm{Tr} \, \Ctl }{\left(\sigma^2 + \frac{r}{t}L \Csup^2\right)^2}.
  \]
  Therefore $\inf_t\left\{\min_l v_l\right\} \geq \frac{ \lambda_1 }{\left(\sigma^2 + k_1\right)^2} $, where
  $\lambda_1= \min_l\left\{ \inf_t \left[ \frac{1}{t}\mathrm{Tr} \, \Crl  \right], \inf_t \left[\frac{1}{t}\mathrm{Tr} \, \Ctl  \right]\right\} >0$
  and $k_1=L \Csup^2 \max\left\{ 1, \inf_t [r/t]\right\}=L\Csup\lambda_0 < +\infty$.
  Noting $k_0=\frac{\lambda_1}{\lambda_0}>0$
  we can now conclude about statement \eqref{it:rspM} of the lemma:
  \[
   \sup_t\rho(\M(\T,\tilde\T))
   \leq 1 - \frac{\inf_t (\min_l {\bf v}_l)}{\sup_t (\max_l {\bf u}_l)}
   \leq 1 - \frac{k_0 \sigma^2}{(\sigma^2 + k_1)^2}.
  \]
  
  As for statement \eqref{it:rspAAt} of the lemma, we note that $\big| \M(\T,\tilde\T) - \lambda \I_{2L} \big| = \left| \sigma^4 \tilde\A(\tilde\T)\A(\T) - \lambda^2 \I_{L} \right|$.
  Hence
  $\rho(\sigma^4\tilde{\bf A}(\tilde{\bf T}){\bf A}({\bf T}))=(\rho(\M(\T,\tilde\T)))^2 \leq \left(  1 - \frac{k_0 \sigma^2}{(\sigma^2 + k_1)^2} \right)^2 <1$.

  Concerning statement \eqref{it:tnorm}, the proof is the same as in \cite[Lemma 5.2]{hachem2008clt}.
  Nonetheless we provide it here for the sake of completeness.
  As $\rho(\M(\T,\tilde\T))<1$,
  the series $\sum_{k\in\mathbb{N}}\M(\T,\tilde\T)^k$ converges,
  matrix $\I_{2L}-\M(\T,\tilde\T)$ is invertible and its inverse can be written as
  $\left(\I_{2L}-\M(\T,\tilde\T)\right)^{-1}=\sum_{k\in\mathbb{N}}\M(\T,\tilde\T)^k$.
  Therefore the entries of $\left(\I_{2L}-\M(\T,\tilde\T)\right)^{-1}$ are non-negative and
  \[
   {\bf u}_k=\sum_{l=1}^{2L} \left[\left(\I_{2L}-\M(\T,\tilde\T)\right)^{-1}\right]_{kl} {\bf v}_l \geq \min_l ({\bf v}_l) \sum_{l=1}^{2L} \left[\left(\I_{2L}-\M(\T,\tilde\T)\right)^{-1}\right]_{kl}.
  \]
  Hence
  $\max_k \sum_{l=1}^L \left[\left(\I_{2L}-\M(\T,\tilde\T)\right)^{-1}\right]_{kl} \leq \frac{\max_l ({\bf u}_l)}{\min_l ({\bf v}_l)}$.
  As the entries of $\left(\I_{2L}-\M(\T,\tilde\T)\right)^{-1}$ are non-negative, it eventually follows that:
  \[
   \sup_t\left[\tnorm[\left(\I_{2L}-\M(\T,\tilde\T)\right)^{-1}]\right]\leq \frac{\sup_t (\max_l {\bf u}_l)}{\inf_t (\min_l {\bf v}_l)} \leq \frac{(\sigma^2 + k_1)^2}{k_0 \sigma^2}.
  \]

  \end{proof}\medskip
  
  \begin{remark}
   Lemma \ref{lm:rsp} \eqref{it:rspAAt} is used in the proof of Theorem \ref{th:ex-un-algo-cano} for the uniqueness of solutions to \eqref{eq:canonique},
   but we took care not to use any consequences of this uniqueness in the proof above; this proof only requires the existence of solutions to \eqref{eq:canonique}.
  \end{remark}
  
  \begin{remark}
   Unfortunately the assumptions
   $\inf_t\left(\frac{1}{t}\mathrm{Tr\,} \Crl\right)>0$ and $\inf_t\left(\frac{1}{t}\mathrm{Tr}\,\Ctl \right)>0$
   made in Lemma \ref{lm:rsp} cannot be restrained, as
   $\frac{1}{t}\mathrm{Tr}\left(\Crl\T\T\right) \leq \frac{1}{\sigma^4} \left( \frac{1}{t}\mathrm{Tr}\,\Crl \right)$ and similarly
   $\frac{1}{t}\mathrm{Tr}\left(\Ctl\tilde\T\tilde\T\right) \leq \frac{1}{\sigma^4} \left( \frac{1}{t}\mathrm{Tr}\,\Ctl \right)$.
  \end{remark}

  The entries of $\B(\R,\T)$ and $\tilde\B(\tilde\R,\tilde\T)$ respectively converge to the entries of $\A(\T)$ and $\tilde\A(\tilde\T)$, hence
  there exists $t_0$ such that, for $t>t_0$,
  \begin{itemize}
   \item the matrix $\I_{2L}-\N(\R,\T,\tilde\R,\tilde\T)$ is invertible,
   \item $\sup_t\Big[ \tnorm[(\I_{2L}- \N(\R,\T,\tilde\R,\tilde\T))^{-1}] \Big] \leq \frac{2(\sigma^2+k_1)^2}{k_0 \sigma^2}$.
  \end{itemize}
  Then, for $t>t_0$, \eqref{eq:delta-alpha_matrix} yields
  \begin{equation}
    \begin{bmatrix}  \bs\alpha -\bs\delta \\ \bs{\tilde\delta}-\bs{\tilde\alpha} \end{bmatrix}
    = \left( \I_{2L} - \N(\R,\T,\tilde\R,\tilde\T) \right)^{-1}
    \begin{bmatrix}  \bs\varepsilon \\ \bs{0} \end{bmatrix}.
    \label{eq:alpha-delta_inv}
  \end{equation}
  Hence
  $
    \max_l \big\{ \big|\alpha_l-\delta_l \big|, \big|\tilde\alpha_l-\tilde\delta_l \big| \big\}
    \leq \tnorm[( \I_{2L} - \N(\R,\T,\tilde\R,\tilde\T) )^{-1}] \max_k \left|\bs\varepsilon_k\right|
  $,
  and as $\bs\varepsilon_l = \mathrm{Tr} \left( \Crl  \bs\Upsilon \right) =\mathcal{O}\left( \frac{1}{t^2} \right)$ for $l=1, \ldots, L$, we eventually have that
  \begin{equation}
    \tilde\alpha_l-\tilde\delta_l = \mathcal{O}\left( \frac{1}{t^2} \right).
   \label{eq:alpha-delta=O}
  \end{equation}
  Using \eqref{eq:alpha-delta=O} in \eqref{ineq:tr_R-T} completes the proof of Proposition \ref{prop:2nd-appr}.

\section{Integrability of $\mathbb{E}_\H\left[ \mathrm{Tr} \left( \T - \S \right) \right]$ - Proof of Proposition \ref{prop:integrable}}
\label{apx:integrable}
  We first consider $\mathbb{E}_\H\left[ \mathrm{Tr} \left( \R - \S \right) \right]$,
  which is equal to $\mathrm{Tr}\,\bs\Upsilon$ by Proposition~\ref{prop:S-appr}.
  As noted in Remark~\ref{rem:intgbl_trGA} of Appendix \ref{apx:1st_apprx}, we have
  $\left| \frac{1}{t}\mathrm{Tr}(\bs\Upsilon \A) \right| \leq \frac{1}{\sigma^8 t^2} P_0\left( \frac{1}{\sigma^2} \right)$,
  where $P_0$ is a polynomial with real positive coefficients which do not depend on $\sigma^2$ nor on $t$.
  Therefore
  \begin{equation}
    \left| \mathbb{E}_\H\left[ \mathrm{Tr} \left( \R - \S \right) \right] \right|
    \leq \frac{P_0\left( \frac{1}{\sigma^2} \right)}{\sigma^8 t}.
    \label{eq:intgbl_R-S}
  \end{equation}
  
  We now consider $\mathrm{Tr} \left( \R - \T \right)$.
  Similarly to Appendix \ref{apx:alpha-delta}, there exists $t_0$ such that 
  $\I_{2L} - \N(\R,\T,\tilde\R,\tilde\T)$ is invertible
  and such that
  $\tnorm[( \I_{2L} - \N(\R,\T,\tilde\R,\tilde\T) )^{-1}]  \leq \frac{2(\sigma^2+k_1)^2}{k_0 \sigma^2}$,
  where $k_0$ and $k_1$ are given by Lemma \ref{lm:rsp}.
  Then \eqref{eq:delta-alpha_matrix} implies
  \[
    \big|\tilde\alpha_l-\tilde\delta_l \big| \leq \tnorm[( \I_{2L} - \N(\R,\T,\tilde\R,\tilde\T) )^{-1}] \max_k \left|\bs\varepsilon_k\right|
    \leq \frac{2(\sigma^2+k_1)^2}{k_0 \sigma^2} \max_k \left|\bs\varepsilon_k\right|,
  \]
  where $\bs\varepsilon_k = \mathrm{Tr} \left( \Crk  \bs\Upsilon \right)$.
  Besides, Remark \ref{rem:intgbl_trGA} of Appendix~\ref{apx:1st_apprx} ensures that 
  $|\bs\varepsilon_k| \leq \frac{1}{\sigma^8 t^2} P_1\left( \frac{1}{\sigma^2} \right)$,
  where $P_1$ is a polynomial with real positive coefficients which do not depend on $\sigma^2$ nor on $t$.
  Hence,
  \begin{equation}
   \big| \tilde\alpha_l - \tilde\delta_l \big|
   \leq \frac{P_1\left( \frac{1}{\sigma^2} \right)}{\sigma^8 t^2} \frac{2(\sigma^2+k_1)^2}{k_0 \sigma^2} \ \text{ for } t>t_0, l=1,\ldots,L.
    \label{eq:intgbl_alpha-delta_l}
  \end{equation}
  Using \eqref{eq:intgbl_alpha-delta_l} in \eqref{ineq:tr_R-T} with $\A=\I_r$ then gives:
  \begin{equation}
   \left| \mathrm{Tr}\left( \R-\T \right) \right|
     \leq \frac{1}{\sigma^8 t} \, k_2 \left( 1+\frac{k_1}{\sigma^2} \right)^2 P_1\left( \frac{1}{\sigma^2} \right)  \ \text{ for } t>t_0,
    \label{eq:intgbl_R-T}
  \end{equation}
  where $k_2= \frac{2 L \Csup}{k_0} \sup_t\{r/t\}<+\infty$.
  
  Eventually, \eqref{eq:intgbl_R-S} and \eqref{eq:intgbl_R-T} yield
  $\left|\mathbb{E}_\H\left[ \mathrm{Tr} \left( \T - \S \right) \right] \right| \leq  \frac{1}{\sigma^{8} t} P\left( \frac{1}{\sigma^2} \right)$ for $t>t_0$,
  where the coefficients of the polynomial
  $P\left( \frac{1}{\sigma^2} \right)=\left(P_0\left( \frac{1}{\sigma^2} \right) + k_2\left( 1+\frac{k_1}{\sigma^2} \right)^2 P_1\left( \frac{1}{\sigma^2} \right) \right)  $
  are real positive coefficients and do not depend on $\sigma^2$ nor on $t$.
  This completes the proof of Proposition \ref{prop:integrable}.

\section{Differentiability of $\Q  \mapsto \bs\delta(\Q)$, $\Q \mapsto \bs{\tilde\delta}(\Q)$ and $\Q \mapsto \overline{I}(\Q)$ - Proof of Proposition \ref{prop:diff_Ib}}
\label{proof:diff}

We prove in this section that, for ${\bf Q}, {\bf P} \in \mathcal{C}_1$, functions $\bs{\delta}$ and $\bs{\tilde\delta}$ are G\^ateaux differentiable at point ${\bf Q}$ in the direction ${\bf P}-{\bf Q}$, where $\bs{\delta}, \bs{\tilde\delta}$ are defined as the solutions of system (\ref{eq:canonique}).
The proof is based on the implicit function theorem.

Let ${\bf P}, {\bf Q} \in \mathcal{C}_1$. We introduce the function $\Gamma:\mathbb{R}_+^L \times \mathbb{R}_+^L \times [0,1] \rightarrow \mathbb{R}^{2L}$ defined by
\[
 \Gamma(\bs{\delta},\bs{\tilde\delta},\lambda)= \left[
 \begin{array}{c}
  \bs{\delta}-f(\bs{\tilde\delta})
  \\ 
  \bs{\tilde\delta}-\tilde f(\bs{\delta},{\bf Q}+\lambda({\bf P}-{\bf Q}))
 \end{array}
 \right],
\]
with $f(\bs{\tilde\delta})=\left[f_1(\bs{\tilde\delta}),\, \ldots,\, f_L(\bs{\tilde\delta})\right]^T$ and $\tilde f(\bs{\delta},{\bf Q})=\left[\tilde f_1(\bs{\delta},{\bf Q}),\, \ldots,\, \tilde f_L(\bs{\delta},{\bf Q})\right]^T$, where the $f_l$ and the $\tilde f_l$ are defined by (\ref{eq:canoniquebis}).
Note that $\bs{\delta}({\bf Q}+\lambda({\bf P}-{\bf Q}))$ and $\tilde{\bs{\delta}}({\bf Q}+\lambda({\bf P}-{\bf Q}))$ are defined by $\Gamma(\bs{\delta},\bs{\tilde\delta},\lambda)=0$.
We want to apply the implicit theorem on a neighbourhood of $\lambda=0$; this requires the differentiability of $\Gamma$ on this neighbourhood, and the invertibility of the partial Jacobian $ D_{(\bs{\delta},\bs{\tilde\delta})}(\Gamma(\bs{\delta},\bs{\tilde\delta},\lambda))$ at point $\lambda=0$.

We first note that
$f_l:\bs{\tilde\delta}\mapsto\frac{1}{\sigma^2t}\mathrm{Tr}\left[\Crl\left(\I+\sum_k\tilde\delta_k\Crk\right)^{-1}\right]$
is clearly continuously differentiable on $\mathbb{R}_+^L$. 
Concerning $\tilde{f}_l$, we first need to use the matrix equality $({\bf I}+{\bf AB})^{-1}{\bf B}={\bf B}({\bf I}+{\bf BA})^{-1}$, with ${\bf A}={\bf Q}^{1/2}$ and ${\bf B}=\tilde{\bf C}{\bf Q}^{1/2}$:
\begin{align}
 \tilde f_l(\bs{\delta},{\bf Q})&=\frac{1}{\sigma^2t}\mathrm{Tr}\left[{\bf Q}^{1/2}\Ctl{\bf Q}^{1/2}\left({\bf I}+{\bf Q}^{1/2}\tilde{\bf C}(\bs{\delta}){\bf Q}^{1/2}\right)^{-1}\right] \notag
 \\
 &=\frac{1}{\sigma^2t}\mathrm{Tr}\left[\Ctl{\bf Q}({\bf I}+\tilde{\bf C}(\bs{\delta}){\bf Q})^{-1}\right].
 \label{eq:ftilde2}
\end{align}
Recall that $\tilde\C(\bs\delta)=\sum_k\delta_k\Ctk$.
Function $(\bs{\delta},\lambda)\mapsto \tilde f(\bs{\delta},{\bf Q}+\lambda({\bf P}-{\bf Q}))$ is therefore clearly continuously differentiable on $\mathbb{R}^{+L} \times [0,1]$.
Nevertheless, as we want to use the implicit theorem for $\lambda=0$, we need to enlarge the continuous differentiability on an open set including $\lambda=0$.
Note that for $\lambda<0$, ${\bf Q}+\lambda({\bf P}-{\bf Q})$ might have negative eigenvalues.
Yet, $\det\left[{\bf I}+\tilde{\bf C}(\bs{\delta})({\bf Q}+\lambda({\bf P}-{\bf Q}))\right]>0$ for $\bs{\delta}=\bs{\delta}({\bf Q})$ and $\lambda=0$.
Therefore it exists a neighbourhood $V$ of $(\bs{\delta}({\bf Q}),0)$ on which $\det\left[{\bf I}+\tilde{\bf C}(\bs{\delta})({\bf Q}+\lambda({\bf P}-{\bf Q}))\right]>0$.
Defining $\tilde{f}_l$ by (\ref{eq:ftilde2}), the functions $(\bs{\delta},\lambda)\mapsto \tilde f_l(\bs{\delta},{\bf Q}+\lambda({\bf P}-{\bf Q}))$ are continuously differentiable on $V$.
Hence, $\Gamma(\bs{\delta},\tilde{\bs{\delta}},\lambda)$ is continuously differentiable on $\mathbb{R}^L \times V$.

We still have to check that the partial Jacobian $ D_{(\bs{\delta},\bs{\tilde\delta})}(\Gamma(\bs{\delta},\bs{\tilde\delta},\lambda))$ is invertible at the point $\lambda=0$.
\[
 D_{(\bs{\delta},\bs{\tilde\delta})}\Gamma_{(\bs{\delta},\bs{\tilde\delta},0)}
  =\begin{bmatrix} {\bf I}-D_{\bs{\delta}}f_{(\tilde{\bs{\delta}})} & -D_{\tilde{\bs{\delta}}}f_{(\tilde{\bs{\delta}})}
      \\ -D_{\bs{\delta}}\tilde f_{({\bs{\delta}},0)} & {\bf I}-D_{\tilde{\bs{\delta}}}\tilde f_{(\bs{\delta},0)} \end{bmatrix}
  =\begin{bmatrix} {\bf I} & -\sigma^2{\bf A}({\bf T}) \\ -\sigma^2\tilde{\bf A}(\tilde{\bf T}) & {\bf I} \end{bmatrix}
  =\M(\T,\tilde\T),
\]
where
$\A_{kl}(\T)=\frac{1}{t}\mathrm{Tr}(\Crk\T\Crl\T)$
and
$\tilde\A_{kl}(\tilde\T)=\frac{1}{t}\mathrm{Tr}(\Q^{1/2}\Ctk\Q^{1/2}\tilde\T\Q^{1/2}\Ctl\Q^{1/2}\tilde\T)$,
and where
$\T=\T(\tilde{\bs{\delta}}(\Q))$ and  $\tilde\T=\tilde\T(\bs{\delta}(\Q))$ are defined by (\ref{eq:canoniqueter}).
Matrices $\A(\T)$, $\tilde\A(\tilde\T)$ and $\M(\T,\tilde\T)$ correspond to those defind in Lemma \ref{lm:rsp}, but in which $\Ctl$ is replaced by $\Q^{1/2}\Ctl \Q^{1/2}$.
Lemma \ref{lm:rsp}, \eqref{it:rspM} therefore gives the invertibility of $ D_{(\bs{\delta},\bs{\tilde\delta})}\Gamma$ at point $\lambda=0$.

We now are in position to apply the implicit function theorem, which asserts that functions $\lambda\mapsto\bs{\delta}({\bf Q}+\lambda({\bf P}-{\bf Q}))$ and $\lambda\mapsto\tilde{\bs{\delta}}({\bf Q}+\lambda({\bf P}-{\bf Q}))$ are continuously differentiable on a neighbourhood of $0$. Hence, $\bs{\delta}$ and $\bs{\tilde\delta}$ are G\^ateaux differentiable at point ${\bf Q}$ in the direction ${\bf P}-{\bf Q}$.
As
$\overline{I}(\Q) = \log \left| \I + \sum_l \dtl(\Q) \Crl \right| +\log \left|{\bf I} + \Q \left( \sum_l \drl(\Q) \Ctl \right) \right| -\sigma^{2} t \left( \sum_l  \drl(\Q)  \dtl(\Q) \right)$
it is clear that $\Q \mapsto \overline{I}(\Q)$ is as well G\^ateaux differentiable at point ${\bf Q}$ in the direction ${\bf P}-{\bf Q}$.

\bibliographystyle{IEEEtran}

\bibliography{IEEEabrv,bibMarne}

\end{document}